%% file: arxiv.tex
\documentclass[12pt,a4paper]{article}

\usepackage[
    a4paper,
    left=20mm,
    right=20mm,
    top=30mm,
    bottom=20mm
]{geometry}

\usepackage[english]{babel}
\usepackage{microtype}
\usepackage{setspace}
\usepackage{hyphenat}
\usepackage{ragged2e}
\usepackage{enumitem}
\usepackage{multicol}

\usepackage{amsmath}
\usepackage{amssymb}
\usepackage{amsbsy}
\usepackage{mathtools}
\usepackage{bm}
\usepackage{dsfont}
\usepackage{mathrsfs}
\usepackage{xfrac}
\usepackage{nicefrac}
\usepackage{siunitx}

\usepackage{graphicx}
\usepackage{svg}
\usepackage{pdfpages}
\usepackage{float}
\usepackage{placeins}
\usepackage{rotating}
\usepackage{morefloats}

\usepackage[dvipsnames,svgnames,table]{xcolor}
\usepackage[font=footnotesize]{caption}
\usepackage{subcaption}

\graphicspath{
    {main_figures/}
    {supplementary_figures/}
}

\usepackage{array}
\usepackage{tabularx}
\usepackage{longtable}
\usepackage{supertabular}
\usepackage{booktabs}
\usepackage{multirow}
\usepackage{makecell}
\usepackage{dcolumn}
\usepackage{threeparttable}
\usepackage{arydshln}
\usepackage{adjustbox}
\usepackage{colortbl}

\newcolumntype{L}[1]{%
    >{\raggedright\arraybackslash}p{#1}
}

\usepackage[ruled]{algorithm2e}
\usepackage{algorithmic}
\usepackage{fvextra}
\usepackage{comment}
\usepackage{appendix}
\usepackage{titlesec}
\usepackage{authblk}

\usepackage[super,comma]{natbib}
\usepackage{hyperref}
\usepackage{cleveref}

\newcommand{\beginsupplement}{%
    \clearpage

    \setcounter{section}{0}
    \renewcommand{\thesection}{\arabic{section}}

    \setcounter{subsection}{0}
    \renewcommand{\thesubsection}{%
        \thesection.\arabic{subsection}%
    }

    \setcounter{equation}{0}
    \renewcommand{\theequation}{S\arabic{equation}}

    \setcounter{table}{0}
    \renewcommand{\thetable}{\arabic{table}}

    \setcounter{figure}{0}
    \renewcommand{\thefigure}{\arabic{figure}}

    \renewcommand{\theHsection}{supp.\arabic{section}}
    \renewcommand{\theHfigure}{supp.\arabic{figure}}
    \renewcommand{\theHtable}{supp.\arabic{table}}
    \renewcommand{\theHequation}{supp.\arabic{equation}}

    \captionsetup[figure]{name=Supplementary Figure}
    \captionsetup[table]{name=Supplementary Table}

    \titleformat{\section}
        {\normalfont\large\bfseries}
        {Supplementary Note~\thesection}
        {1em}
        {}

    \titleformat{\subsection}
        {\normalfont\normalsize\bfseries}
        {\thesubsection}
        {1em}
        {}
}

\title{%
From population norms to personalized trajectories:
interpretable Bayesian forecasting for cognitive decline
}

\author[1,*]{Maria Sahakyan}
\author[1,2]{Brita Elvevåg}

\affil[1]{%
Department of Clinical Medicine,
UiT -- The Arctic University of Norway,
Tromsø, Norway
}

\affil[2]{%
The Norwegian Center for Clinical Artificial Intelligence,
University Hospital of North Norway
}

\affil[*]{%
Corresponding author:
\href{mailto:maria.sahakyan@uit.no}{maria.sahakyan@uit.no}
}

\date{}

\begin{document}

\maketitle
\vspace{-2em}
\begin{abstract}
\noindent \textbf{Background:} The timing and rate of cognitive decline vary substantially between individuals, limiting the ability of fixed population-level thresholds to determine whether a new observation represents meaningful change for an individual. We developed Personalized Risk Inference via Sequential Monitoring (PRISM), an interpretable framework for individualized longitudinal forecasting of cognitive decline. 

\noindent \textbf{Methods:} PRISM estimates a personalized cognitive baseline from routinely collected demographic, health, and functional variables using an Explainable Boosting Machine, then updates this expectation through Bayesian inference with temporal decay as repeated cognitive scores accrue. Decline is evaluated relative to an age-adjusted personal anchor, with uncertainty quantified through posterior probabilities. We evaluated PRISM in 30,664 adults from the Health and Retirement Study and externally validated it in 1,866 Alzheimer’s Disease Neuroimaging Initiative participants. Forecasting performance was compared with demographic-norm and cumulative-average baselines, and discrimination with a linear mixed-effects model. 

\noindent \textbf{Results:} PRISM identified emerging decline before study-defined cognitive worsening in 31\% of sustained decliners in the Health and Retirement Study and 41\% in the Alzheimer’s Disease Neuroimaging Initiative, with median lead times of 6 and 2 years, respectively. By the time of worsening, 68\% and 56\% had been identified. PRISM also achieved lower forecasting error than demographic-norm and cumulative-average baselines and distinguished worsening from stable trajectories better than a standard linear mixed-effects model, particularly early in follow-up. 

\noindent \textbf{Conclusions:} PRISM enables earlier, interpretable, uncertainty-aware detection of cognitive decline relative to each individual’s expected trajectory using routinely collected data. It may support closer monitoring and timely assessment when personal longitudinal history is still limited. 

\noindent \textbf{Plain language summary:} This study explored whether cognitive decline could be detected earlier by comparing a person’s test scores with their own expected pattern of performance rather than a single population threshold. Using cognitive data from more than 32,000 older adults across two independent studies, we developed a framework that sets each person’s expected level of performance and adjusts it as new test scores become available. The framework detected emerging cognitive decline several years before it became clinically apparent and forecasted future cognitive scores more accurately than standard methods. These findings suggest that comparing each person's results with their own expected pattern, rather than general threshold, could help clinicians spot early warning signs of cognitive decline earlier, supporting timely care and planning. 
\end{abstract}

\section*{Introduction}

\noindent In aging and neurodegenerative conditions, cognitive decline often progresses gradually~\cite{tahami2022alzheimer,wallon2025challenging} with pathological changes typically starting many years before noticeable clinical symptoms emerge~\cite{rasmussen2019alzheimer,sperling2011toward}. This extended preclinical window is critically important: emerging disease-modifying treatments may be most effective before significant cognitive impairment develops, while timely identification can support preventative lifestyle interventions, personalized care planning, and optimized clinical trial enrollment~\cite{reetz2026best}. Yet, early signs of decline are often subtle, and trajectories vary substantially across individuals and dementia subtypes, with different subtypes showing distinct early and prominent cognitive symptoms. This heterogeneity complicates timely detection and highlights the limitations of one-size-fits-all prognostic strategies~\cite{koksalmis2025artificial}. Current approaches for detecting cognitive decline, in clinical screening and large-scale epidemiological studies, often rely on fixed or broadly adjusted population-level cutoffs: individuals are  classified as impaired when their performance on cognitive assessments falls below a predetermined threshold, often defined as approximately two standard deviations below the normative population mean. These thresholds create two fundamental problems. First, they ignore substantial variation in cognitive reserve and prior cognitive functioning. Since individuals begins from different baseline levels of ability, population-average benchmarks may fail to detect meaningful within-person decline~\cite{tucker2009cognitive,just2026challenges}. This reliance on fixed norms implicitly treats population-level variations as a proxy  for individual change over time, an assumption that is difficult to sustain in cognitive aging, given the substantial heterogeneity in trajectories~\cite{voelkle2014toward,fisher2018lack}. Second, population cutoffs are inherently retrospective: decline is identified only after an individual's performance falls below an absolute threshold, meaning that detection often occurs after the optimal window for intervention has narrowed~\cite{mattke2023expanding}.

Several methodological families have attempted to address these limitations, each offering complementary strengths while introducing specific trade-offs.

Linear mixed-effects (LME) models estimate individual-level trajectories through random intercepts and slopes, and have been widely used to model longitudinal decline~\cite{murphy2022accessible,rajagopal2024estimating}. Their appeal lies in their interpretability and ability to separate population-level trends from individual variation. Empirical Bayes and Bayesian formulations further stabilize individual-specific estimates  by assuming that individual trajectories vary around a common population pattern, enabling trajectory estimation in settings with missing or unbalanced  observations~\cite{perez2022evaluating}. However, a reliable estimate of person-specific trajectories generally requires multiple observations per individual. With sparse personal histories, random-effect estimates rely more heavily on the population distribution and may shrink toward population-level patterns, affecting individual-level prediction accuracy and bias~\cite{robinson1991blup,li2022bias}. A further limitation is that standard linear mixed models may produce misleading inferences when applied directly to bounded psychometric cognitive tests, particularly when measurement nonlinearity and ceiling or floor effects are ignored~\cite{proust2011misuse}. In addition, conventional LME models are often specified using relatively simple parametric forms, such as linear or low-order time effects. Although these models are interpretable and well suited for estimating average longitudinal trends, such specifications may be less flexible for capturing nonlinear or rapidly evolving decline, especially in neurodegenerative settings where cognitive and clinical trajectories can be nonlinear~\cite{guerrero2016instantiated,samtani2012improved}.

More flexible longitudinal frameworks, including state-space models, survival models, Bayesian joint models, and nonlinear disease-progression models, extend standard LME formulations by linking temporal trajectories with event risk or by modeling nonlinear progression patterns~\cite{rizopoulos2011dynamic,chua2022state,brookmeyer2018forecasting,cauchi2024individualized}. These approaches can update the risk over time and provide principled uncertainty estimates, making them valuable for modeling disease progression. In Alzheimer's disease research, recent work has also shown that nonlinear disease-progression models and recurrent neural networks can forecast neuropsychological and biomarker progression from patient data and support prognostic enrichment in clinical trials, including the selection of participants likely to progress during the trial period~\cite{maheux2023forecasting}. However, these models often require richer longitudinal information, careful handling of missing or irregular measurements, and more complex model specification or pre-processing, which can make implementation and interpretation challenging~\cite{chen2025reflections}. As a result, performance may remain sensitive to missingness, irregular follow-up, and sparse individual histories. These approaches underscore the value of personalized forecasting in neurodegenerative diseases, while also motivating complementary frameworks for interpretable, visit-by-visit monitoring in sparse cognitive assessment settings.

In parallel, machine learning approaches have shown strong predictive performance in modeling cognitive decline. Cross-sectional models can support risk stratification from baseline or single-visit measurements~\cite{stamate2020applying}; however, because they rely on static information, they do not explicitly model within-person temporal dynamics. Longitudinal machine learning models have therefore been proposed to incorporate repeated measurements and capture evolving cognitive and functional change over time~\cite{ding2023prediction,nguyen2018modeling,liu2025longitudinal}. These models can capture complex temporal patterns across repeated assessments, and deep learning approaches have shown strong performance in Alzheimer's disease classification and prognostic prediction~\cite{jo2019deep}. However, such models are often data-intensive and may lack interpretability and clinical transparency~\cite{mourby2021transparency}. 

Beyond routinely collected cognitive and clinical variables, modality-specific approaches may provide additional sensitivity to early cognitive changes. Speech-based methods are relatively scalable and may be sensitive to early cognitive changes~\cite{sharafeldeen2025machine}, but their clinical use still requires rigorous evaluation, standardized acquisition procedures, and careful quality control~\cite{robin2020evaluation}. EEG and neuroimaging approaches can provide additional brain-based markers of cognitive impairment~\cite{meghdadi2021resting,wang2020neuroimaging,mcconathy2015imaging}, but require specialized acquisition infrastructure that may limit scalability in routine clinical settings. Thus, strong prediction alone does not fully address the need for interpretable, uncertainty-aware monitoring that can operate when longitudinal history is sparse and only routinely collected data are available.

Despite these advances, three challenges continue to limit early personalized detection of cognitive decline. First, longitudinal monitoring faces a \textit{cold start problem} when only one or two observations are available, models have limited information from which to estimate a reliable personalized baseline or trajectory~\cite{schein2002coldstart,singh2024systematic}. In practice, this early-data limitation is often addressed  using population-level normative references, where an individual's performance is interpreted relative to demographically matched averages~\cite{Sachs17112022,sosa2009population}. However, such references provide limited sensitivity to within-person decline, particularly for individuals whose premorbid performance differs substantially from the population mean. Second, effective monitoring requires moving beyond static reference points toward probabilistic predictions that update as new observations become available. Recent work has emphasized the importance of hybrid baselines that combine stable reference anchors with personalized longitudinal adaptation for interpreting behavioral change over time~\cite{just2026challenges}. Although Bayesian and joint modeling frameworks provide principled uncertainty quantification and dynamic individualized prediction, separating gradual pathological decline from normal age-related variation remains challenging because cognitive changes may unfold slowly, and the boundary between healthy aging and early neurodegenerative disease is often difficult to define~\cite{FJELL201420,harada2013normal}. Third, a persistent trade-off remains between predictive performance and interpretability~\cite{caruana2012intelligible}. Many high-performing models lack the transparency needed for clinical use~\cite{rudin2019stop}, while interpretable approaches are often not designed for sequential updating, uncertainty-aware monitoring, or personalized trajectory assessment. 

These challenges suggest that early detection of cognitive decline requires a framework that can initialize personalized predictions before a rich individual history is available, update those predictions as new observations accumulate, and remain interpretable at each stage. 

The components needed for such a framework exist, but largely in isolation. Interpretable models such as Explainable Boosting Machines (EBMs)~\cite{nori2019interpretml} can capture non-linear relationships while providing transparent feature-level contributions. When applied to population-level data, they can generate personalized expected baselines from demographic and clinical characteristics~\cite{caruana2015readmission,mahamadou2025interpretable}. In parallel, Bayesian longitudinal frameworks enable predictions to be updated sequentially as new observations are collected, providing uncertainty-aware estimates that evolve over time~\cite{rizopoulos2011dynamic}. However, these approaches are rarely integrated: interpretable models are often applied in static settings, while Bayesian frameworks are not typically anchored to personalized, interpretable baselines. Moreover, many longitudinal prediction approaches emphasize group-level timing, aggregate predictive performance, or event risk, rather than explicitly quantifying individual-level detection relative to clinical transition. Integrating these perspectives would allow a population-informed baseline to be  mapped to each individual and then refined through Bayesian updating as new cognitive observations are acquired, yielding a transparent and dynamically updated estimate of cognitive trajectory and decline probability. 

We present PRISM (Personalized Risk Inference via Sequential Monitoring), an interpretable three-stage framework for cognitive monitoring and forecasting. PRISM uses an Explainable Boosting Machine (EBM) to estimate a personalized cognitive baseline from demographic and health characteristics, updates this baseline through Bayesian inference as new test scores are observed, and estimates the probability of decline from the individual's evolving trajectory. These continuous probabilistic outputs can then be translated into clinically meaningful detection rules for identifying decline beyond that expected from healthy aging. 

This integrated approach frames cognitive decline detection as a prospective monitoring and forecasting problem rather than a fixed horizon classification task, aligning with recent perspectives that conceptualize longitudinal behavioral interpretation as a forecasting rather than static classification~\cite{just2026challenges}. It also enables explicit quantification of when the decline is detected relative to clinical transition. Existing approaches typically report decline retrospectively relative to a later diagnosis or evaluate discrimination within fixed prediction horizons. They generally do not prospectively assess the proportion of individuals before their observed classification change through visit-by-visit monitoring. 

We evaluated PRISM in two longitudinal cohorts, the Health and Retirement Study (HRS) and Alzheimer's Disease Neuroimaging Initiative (ADNI), to test whether personalized Bayesian monitoring could detect emerging cognitive decline before study-defined cognitive worsening. Since decline may unfold through heterogeneous trajectories, including gradual deterioration, abrupt worsening, and transient fluctuations, PRISM uses complementary detection rules rather than a single fixed criterion. We assessed both forecasting performance and the timing of prospective detection relative to clinical transition. 

\section*{Results}

PRISM is a three-stage monitoring framework that estimates a personalized EBM baseline, refines this estimate using Bayesian updating as longitudinal observations accumulate, and estimates the posterior probability of decline from the longitudinal slope (Figure~\ref{fig:PRISM_framework}; full details in Methods and Supplementary Note~\ref{note:PRISM}).

\subsection*{Longitudinal forecasting and Bayesian adaptation in HRS}

We first evaluated whether PRISM could produce stable personalized longitudinal forecasts in the HRS test cohort ($N=4{,}600$; 1996--2020). As observations accumulated, predictions transitioned from population-informed EBM prior toward personalized longitudinal trajectories.

\begin{figure}[!ht]
    \centering
    \includegraphics[width=\textwidth]{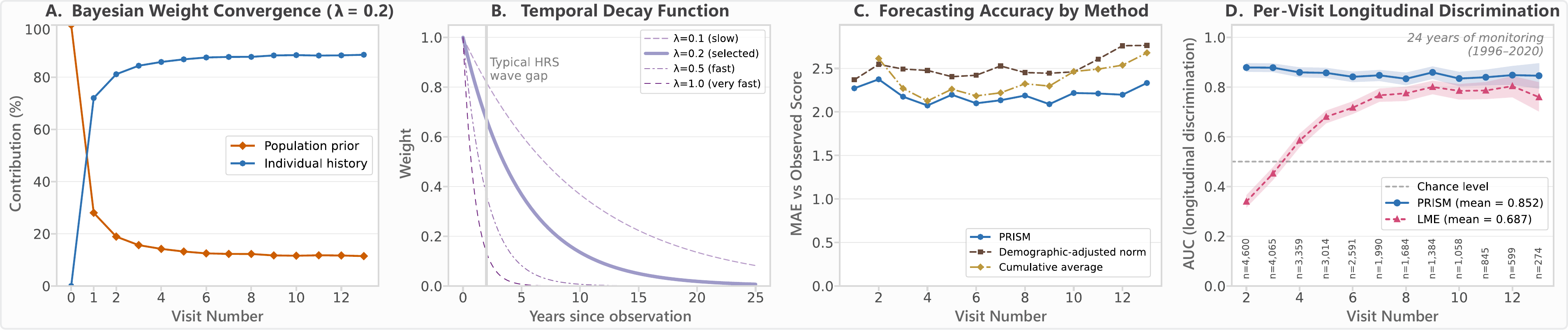}
    \caption{\textbf{Longitudinal behavior and forecasting performance of the PRISM framework in the Health and Retirement Study (HRS) test set ($N = 4{,}600$; 1996--2020).}
    \textbf{(A) Bayesian adaptation from population prior to individual history.}
     Relative contribution of the EBM-derived population prior and individualized longitudinal history to the posterior estimate across visits. Individual history rapidly becomes dominant after the first observed visit.
    \textbf{(B) Temporal decay weighting ($\lambda$).}
    Exponential decay functions showing how prior observations are weighted as a function of elapsed time for different values of the decay parameter. The selected value, $\lambda = 0.2$, was optimized on the validation set. The vertical line indicates the typical 2-year interval between HRS assessment waves. At the selected decay rate, observations retain approximately 67\% of their weight across one inter-wave interval,
    $w(t) = e^{-\lambda t}$ and $w(2) = e^{-0.2 \times 2} \approx 0.67$,
    allowing recent assessments to dominate forecasting while preserving informative longitudinal history.
    \textbf{(C) Forecasting accuracy across visits.}
    One-step-ahead forecasting accuracy across visits, measured by mean absolute error (MAE), comparing PRISM with a demographic-adjusted norm baseline and a cumulative individual-average baseline.
    \textbf{(D) Per-visit longitudinal predictive performance.}
    Area under the receiver operating characteristic curve (AUC) for distinguishing individuals with worsening cognitive status from those who remain stable, evaluated independently at each visit. PRISM scores were based on posterior decline probability, whereas LME scores were based on participant-specific random slopes. Shaded regions indicate 95\% bootstrap confidence intervals.}
    \label{fig:prism_longitudinal_performance}
\end{figure}

Analysis of Bayesian weight dynamics demonstrated a rapid shift from population-informed priors to individualized estimation (Figure~\ref {fig:prism_longitudinal_performance}A). Following a single observed visit, individual longitudinal information contributed approximately 72\% of the posterior estimate, indicating a marked reduction in uncertainty once participant-specific cognitive evidence became available. Subsequently, the individualized longitudinal history remained the dominant contributor throughout follow-up, while the population-informed prior continued to provide a smaller stabilizing contribution through Bayesian updating and temporal weighting of earlier observations. 
The temporal decay parameter optimized on the validation set ($\lambda=0.2$) maintained considerable influence on recent observations while progressively reducing the impact of older assessments (Figure~\ref {fig:prism_longitudinal_performance}B). At the typical two-year interval between HRS waves, previous observations preserved approximately 67\% of their weight, allowing the framework to remain sensitive to recent cognitive changes while still retaining informative longitudinal history. 

To evaluate forecasting accuracy, PRISM's one-step-ahead predictions were compared against two reference baselines representing different levels of personalization---a static demographic-adjusted population norm and a cumulative individual average---using mean absolute error (MAE; Supplementary Note~\ref{note:evaluation_metrics}). Across all visits, PRISM achieved lower one-step-ahead forecasting errors than both the static demographic-adjusted population norm and the cumulative individual average baselines (Figure~\ref {fig:prism_longitudinal_performance}C). Differences between these approaches were most pronounced in early follow-up, when a limited longitudinal history increased reliance on the individualized EBM-derived prior. Although forecasting improvements were modest at some visits, PRISM also provides outputs that the reference baselines do not: individualized uncertainty-aware decline probabilities, personalized anchors, temporal weighting of prior observations, and prospective forecasts from the earliest visits, before sufficient longitudinal history is available.

To contextualize PRISM's visit-level discrimination performance against an established repeated-measures approach, we additionally compared it with a standard linear mixed-effects model (LME) with participant-specific random intercepts and slopes for time, and fixed effects for age, sex, and education. PRISM showed substantially stronger per-visit discrimination between individuals with worsening versus stable cognitive status than the LME-derived participant-specific slope signal, particularly during early follow-up when mixed-effects estimates are heavily regularized toward the population trajectory (Figure~\ref {fig:prism_longitudinal_performance}D; Supplementary Note~\ref{note:evaluation_metrics}). This difference likely reflects the difficulty of estimating reliable participant-specific slopes when personal longitudinal history is sparse and the pool of individual trajectories is regularized toward a single population mean slope. These results suggest that estimating longitudinal trajectories and identifying emerging decline are related but not equivalent objectives.

\subsection*{Early detection of cognitive decline}

\begin{figure}[!ht]
    \centering
    \includegraphics[width=\textwidth]{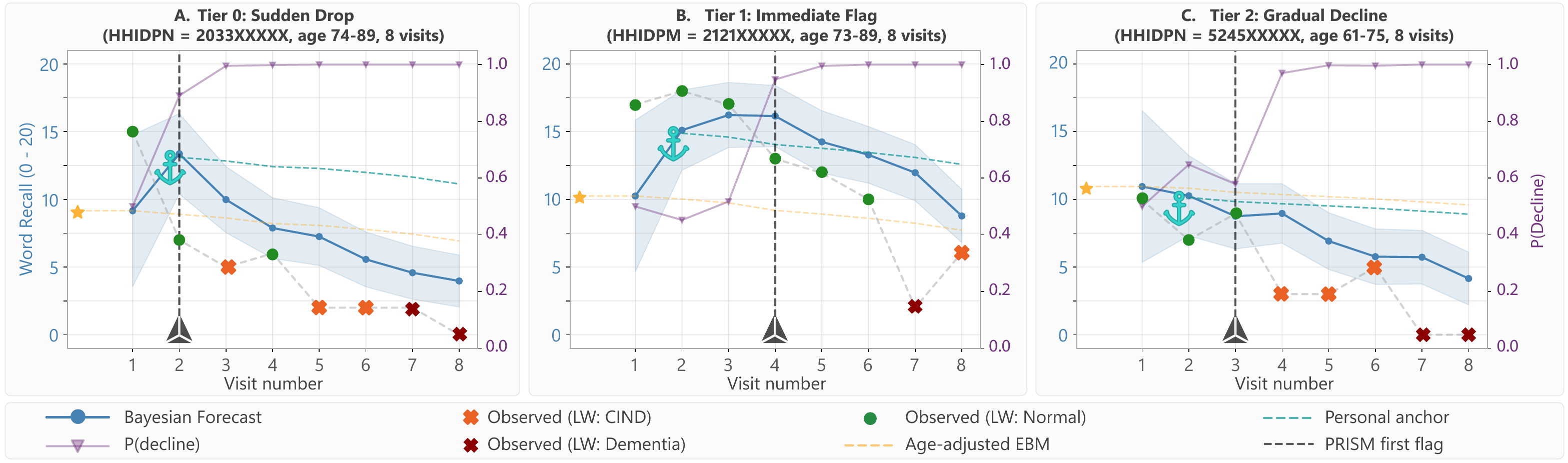}
    \caption{\textbf{Representative cognitive trajectories from the HRS test set illustrating the three PRISM decline-detection tiers.}
    Each panel shows one individual's longitudinal word-recall trajectory together with the corresponding PRISM one-step-ahead Bayesian forecast. Observed visit scores are marked according to the Langa--Weir cognitive classification at that visit: Normal cognition, CIND/MCI, or dementia. The Bayesian forecast is shown as a solid line, with shaded regions indicating the 95\% credible interval. The star marks the initial EBM-derived baseline estimate, the dashed EBM line shows the age-adjusted population-informed reference over time, and the dashed personal-anchor line represents the individualized reference used for below-baseline detection after longitudinal information becomes available. The secondary y-axis shows the posterior probability of decline, $P(\mathrm{decline})$. The vertical dashed line marks the first visit at which PRISM flagged cognitive decline. Panels show representative examples of \textbf{(A)} Tier~0, capturing abrupt severe drops; \textbf{(B)} Tier~1, capturing immediate high-confidence decline; and \textbf{(C)} Tier~2, capturing gradual decline requiring sustained longitudinal evidence.}
    \label{fig:prism_detection_tiers}
\end{figure}

Representative HRS test-set trajectories showed that PRISM identified distinct longitudinal patterns of cognitive decline before participants transitioned to a worse Langa--Weir cognitive category (Figure~\ref{fig:prism_detection_tiers}). The three-tier detection rule captured abrupt severe drops, immediate high-confidence decline, and slower gradual deterioration, illustrating how the framework adapts to different temporal profiles of cognitive change rather than relying on a single fixed threshold.

Across the three examples, PRISM combined two complementary sources of evidence: the posterior probability that an individual's trajectory was declining, $P(\mathrm{decline})$, and evidence that cognitive performance had fallen below an age-adjusted personal anchor. This joint criterion prevented isolated increases in $P(\mathrm{decline})$ from triggering a flag unless the corresponding cognitive level was also sufficiently below the individualized reference. At the first observed visit, the EBM-derived baseline served as the initial reference because no individualized longitudinal history was yet available. From the second visit onward, this reference was replaced by a personalized anchor derived from the Bayesian posterior after incorporating the first observed score, thereby combining the participant's observed initial performance with the population-informed EBM estimate.

The tiers differed in the strength and persistence of evidence required for detection (Supplementary Note~\ref{note:three_tier_rule}. Tier~0 captured abrupt severe drops, where the posterior estimate fell substantially below the age-adjusted personal anchor together with minimum evidence of a declining trajectory (Figure~\ref{fig:prism_detection_tiers}A). Tier~1 identified cases with strong posterior evidence of decline and below-anchor performance, allowing immediate detection without requiring confirmation over multiple visits (Figure~\ref{fig:prism_detection_tiers}B). Tier~2 captured slower deterioration by requiring sustained below-anchor performance across consecutive visits together with probabilistic evidence of decline (Figure~\ref{fig:prism_detection_tiers}C). In all three representative cases, PRISM flagged decline before the participant transitioned to a worse Langa--Weir cognitive category.

Since the trajectories in Figure~\ref{fig:prism_detection_tiers} show word recall only, whereas the Langa--Weir classification is derived from the full 27-point cognitive score, including word recall, serial subtraction, and backward counting, changes in cognitive-state markers may reflect changes in other cognitive components and therefore may not coincide with abrupt visible changes in the plotted word-recall trajectory.

\begin{figure}[!ht]
    \centering
    \includegraphics[width=\textwidth]{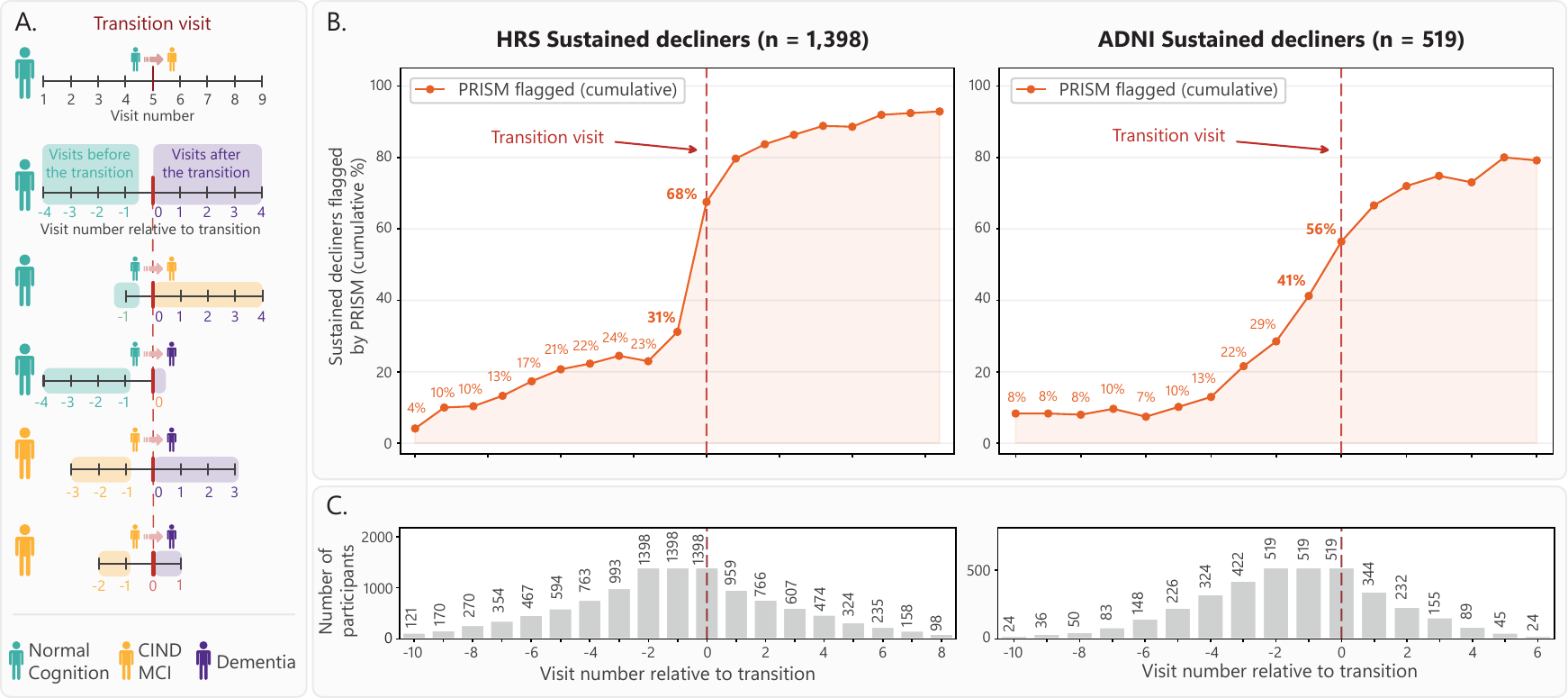}
    \caption{\textbf{Transition-centered alignment of longitudinal trajectories and cumulative early detection by PRISM.}
    \textbf{(A) Transition-centered alignment procedure used for longitudinal analyses.}
    Individual trajectories were aligned relative to the first visit at which worsened cognitive status was observed, defined as the transition visit (visit~0). Negative visit numbers represent assessments preceding the transition, whereas positive visit numbers represent assessments following the transition. Example trajectories illustrate variation across participants in the length of pre-transition history and post-transition follow-up.
    \textbf{(B) Cumulative percentage of sustained decliners flagged by PRISM at each visit relative to transition in the HRS and ADNI cohorts.}
    Curves represent the cumulative proportion of participants who had been flagged by PRISM by or before each relative visit among individuals who ultimately met the sustained-decline endpoint. In HRS, 31\% of sustained decliners had been flagged one visit before transition and 68\% by the transition visit. In ADNI, the corresponding values were 41\% and 56\%, respectively. Dashed vertical lines indicate the transition visit (visit~0).
    \textbf{(C) Number of participants contributing data at each visit relative to transition for the HRS and ADNI cohorts.}
    Sample sizes decrease at more distant pre- and post-transition visits because of variation in the duration of longitudinal follow-up across participants.}
    \label{fig:transition_centered_detection}
\end{figure}

To evaluate how far in advance PRISM detected decline compared to clinically observed cognitive worsening, individual longitudinal trajectories were aligned in time to the first visit at which cognitive status had worsened (``transition visit''; visit 0; Figure~\ref{fig:transition_centered_detection}A). Aligning participants based on their transition point made it possible to compare individuals with varying follow-up durations and different times of transition within a shared temporal frame. In the HRS cohort, 31\% of sustained decliners had already been identified one visit prior to transition, and 68\% had been identified by or at the transition visit (Figure~\ref{fig:transition_centered_detection}B). Importantly, cumulative detection curves increased progressively across pre-transition visits rather than abruptly at the transition point, indicating that PRISM frequently identified elevated risk of decline before cognitive worsening became detectable under categorical cognitive-status definitions.

The participant distributions in Figure~\ref{fig:transition_centered_detection}C show that fewer participants contributed observations at visits further from transition-aligned visit 0, because participants had different lengths of longitudinal follow-up before and after their individual transition visit. Examples of these varying trajectory lengths are illustrated in Figure~\ref{fig:transition_centered_detection}A.

Examination of individual alerting outcomes revealed patterns characteristic of prospective longitudinal monitoring. Among flagged individuals who did not meet the study-defined endpoint during follow-up, some nevertheless showed measurable longitudinal decline, posterior trajectories below their personalized anchors or received their first flag at the final available observation, where later progression could not be assessed. Conversely, transition cases  without prior PRISM alerts, involved short follow-up or abrupt late worsening, leaving limited opportunity for advance detection. These patterns highlight the complexity of evaluating alerting performance in longitudinal settings, where continuous cognitive trajectories do not always align with discrete diagnostic endpoints.

\subsection*{External validation in ADNI}
To evaluate generalizability beyond the HRS cohort and beyond survey-derived cognitive categorization, PRISM was additionally evaluated in the independent ADNI cohort using clinical diagnostic transitions based on comprehensive cognitive and clinical assessment. Despite substantial differences in cohort structure, assessment frequency, and diagnostic procedures, importantly PRISM demonstrated similar early detection behavior in ADNI (Figure~\ref{fig:transition_centered_detection}B).
Among individuals who met the sustained-decline endpoint, 41\% had already been flagged one visit before transition and 56\% had been flagged by the time of the transition visit. Similar to HRS, detection curves in ADNI showed progressive accumulation of pre-transition flags across visits preceding clinical worsening. The consistency of these temporal detection patterns across HRS and ADNI provides strong evidence that PRISM captures longitudinal patterns of cognitive decline that generalize across both survey-based cognitive categorization and clinically defined diagnostic trajectories. The higher proportion of pre-transition detections in ADNI may partly reflect the shorter interval between assessments relative to HRS, allowing decline-related changes to be captured with finer temporal resolution

\section*{Discussion}

Cognitive decline is an inherently longitudinal, individual-specific, and uncertain process, yet many existing approaches continue to operationalize it using static population thresholds, rigid diagnostic labels, or predictions tied to a fixed time point. Here we introduced PRISM, an interpretable Bayesian framework that reconceptualizes cognitive decline detection as a problem of prospective longitudinal monitoring and forecasting. PRISM combines an Explainable Boosting Machine-derived personalized baseline with uncertainty-aware Bayesian updating, allowing cognitive trajectories to evolve from a population-informed starting point toward increasingly individualized monitoring as longitudinal evidence accumulates. Across two distinct longitudinal cohorts, the Health and Retirement Study (HRS) and the Alzheimer's Disease Neuroimaging Initiative (ADNI), PRISM reliably detected longitudinal evidence of decline before participants first met study-defined criterion for cognitive worsening. Among sustained decliners, PRISM identified 31\% of HRS and 41\% of ADNI participants before their first recorded transition, with cumulative detection increasing to 68\% and 56\%, respectively, by the transition visit. The implications of these findings extend from individualized cognitive decline monitoring to the broader challenge of building explainable, uncertainty-aware forecasting systems for health trajectories.
A central contribution of PRISM is that it provides a reliable and clinically meaningful approach to forecasting cognitive decline. Rather than treating cognitive decline as a static classification problem, PRISM monitors change prospectively and updates each participant's expected trajectory as new observations become available. Its performance relative to static demographic norms and cumulative individual averages suggests that the framework captures longitudinal change that is cognitively meaningful rather than merely statistical. Moreover, PRISM showed stronger discrimination of cognitive worsening than a linear mixed-effects model, indicating that accurate longitudinal score modeling and clinically useful early decline detection are related but distinct objectives. This distinction is important because early cognitive deterioration often unfolds before diagnostic thresholds are crossed, and methods that can detect these emerging changes may support earlier monitoring, further assessment, or timely clinical follow-up. 
PRISM's clinical value also lies in making longitudinal monitoring genuinely personalized. Many longitudinal models require repeated observations before participant-specific estimates become reliable, limiting their usefulness at early visits, when personal cognitive history is sparse. PRISM addresses this cold start problem by using routinely collected baseline characteristics to generate a personalized cognitive expectation before a sufficient individual cognitive history is available. This population-informed prior provides a personalized starting point, but it does not remain fixed: Bayesian updating with temporal decay progressively shifts the framework toward the individual's recent cognitive history as observations accumulate, while retaining a smaller stabilizing contribution from the prior throughout follow-up. This hybrid structure helps avoid two complementary limitations: demographic norms may miss decline in individuals who start well above average, whereas purely history-based approaches can be unstable when observations are sparse or when a single unusually low score occurs. 

This personalized structure is strengthened by PRISM's use of longitudinal anchors rather than fixed cognitive thresholds. Traditional cutoffs use population-level variation as the reference for within-person change, even though individuals differ substantially in baseline cognitive level, cognitive reserve, and expected age-related decline. By comparing current performance against an age-adjusted personalized anchor, PRISM can identify departures from an individual's expected trajectory, including in people whose scores remain within demographically normal range, despite meaningful within-person deterioration. This anchor-based approach makes monitoring person-specific and offers clinical utility beyond what group-level reference standards can provide. 

PRISM also provides explicit measures of uncertainty around decline assessments, a feature that remains uncommon in clinical prediction models. Rather than producing only binary flags or point predictions, the framework estimates P(decline), which reflects both the direction of cognitive change and the strength of evidence supporting that estimate. Conservative probabilities early in follow-up indicate limited longitudinal evidence rather than unwarranted certainty, while additional observations allow PRISM to express increasing confidence in emerging decline patterns. This uncertainty-aware structure helps distinguish cases where decline evidence is strong from cases where continued monitoring is warranted before a confident assessment can be made. More broadly, uncertainty quantification is central to accountable clinical automation: systems that support longitudinal monitoring or clinical decision-making must be able to indicate not only that a risk signal is present, but also how confident that signal is and when the available data remain insufficient for reliable interpretation.

Several limitations and open methodological questions should be considered. First, the HRS reference outcome partly overlaps, creating overlaps with the word recall measure monitored by PRISM, although this concern is mitigated by the ADNI validation using clinician-assigned diagnostic worsening based on broader clinical information and a different memory measure. Second, the current implementation monitors episodic verbal memory only and may miss decline emerging primarily in executive function, language, visuospatial processing, or attention. Third, the temporal decay parameter was optimized globally, although the informativeness of prior observations may vary across individuals, assessment intervals, and clinical contexts. More broadly, prospective longitudinal monitoring raises unresolved evaluation challenges, including how to interpret endpoint-discordant alerts, where a flagged individual shows emerging decline but does not cross a study-defined endpoint within the available follow-up window. It also remains to be tested how PRISM performs in other medical conditions where change may occur more abruptly or over shorter time scales than typical cognitive aging.

Future extensions could examine adaptive or person-specific temporal weighting, incorporate biomarkers, neuroimaging, genetic data, or multimodal sensing, and evaluate PRISM prospectively in real-world clinical settings. These additional could refine individualized forecasting when available, particularly by adding information about underlying neuropathological processes. However, advanced biomarkers are not always routinely collected, and longitudinal biomarker data are even less consistently available. PRISM therefore addresses a complementary forecasting problem: modeling person-specific cognitive trajectories, uncertainty, and decline probability from routinely collected variables and repeated cognitive assessments, while remaining extensible to richer multimodal data when such data are available. More broadly, PRISM illustrates how explainable, uncertainty-aware forecasting can move cognitive decline beyond static thresholds toward adaptive early-warning systems for health trajectories.

In conclusion, PRISM reframes the detection of cognitive decline as a problem of personalized longitudinal forecasting rather than static classification. By combining a population-informed personalized baseline with Bayesian updating, temporal decay, individualized anchors, and probabilistic decline estimation, the framework supports early monitoring even when personal cognitive history is sparse. Across both HRS and ADNI, PRISM identified longitudinal evidence of decline years \textit{before} study-defined cognitive worsening in a substantial subset of individuals who later deteriorated, suggesting that person-specific monitoring can reveal emerging within-person decline well before it appears in categorical transition classifications.

More broadly, PRISM offers a practical framework for forecasting behavioral patterns over time in large, complex, and noisy real-world longitudinal datasets. Its core methodological contribution is the ability to initialize predictions from routinely collected baseline information, then refine these forecasts sequentially as new observations become available, gradually shifting weight from population-level expectations toward individual longitudinal evidence. This formalization may be useful in settings where meaningful cognitive change unfolds over time, varies across individuals, and must be interpreted relative to a person's own trajectory rather than a universal threshold.
PRISM is not intended to replace clinical judgment, but to provide interpretable, uncertainty-aware signals that can support closer monitoring, timely further assessment, and more personalized decision-making. Future work should prospectively evaluate the framework in clinical settings, extend it to multiple cognitive domains and richer data sources, and examine how probabilistic alerts can be integrated into real-world care pathways. More broadly, PRISM illustrates how adaptive longitudinal forecasting can move cognitive decline detection beyond fixed thresholds toward personalized early-warning systems.

\begin{figure}[!ht]
    \centering
    \includegraphics[width=\textwidth]{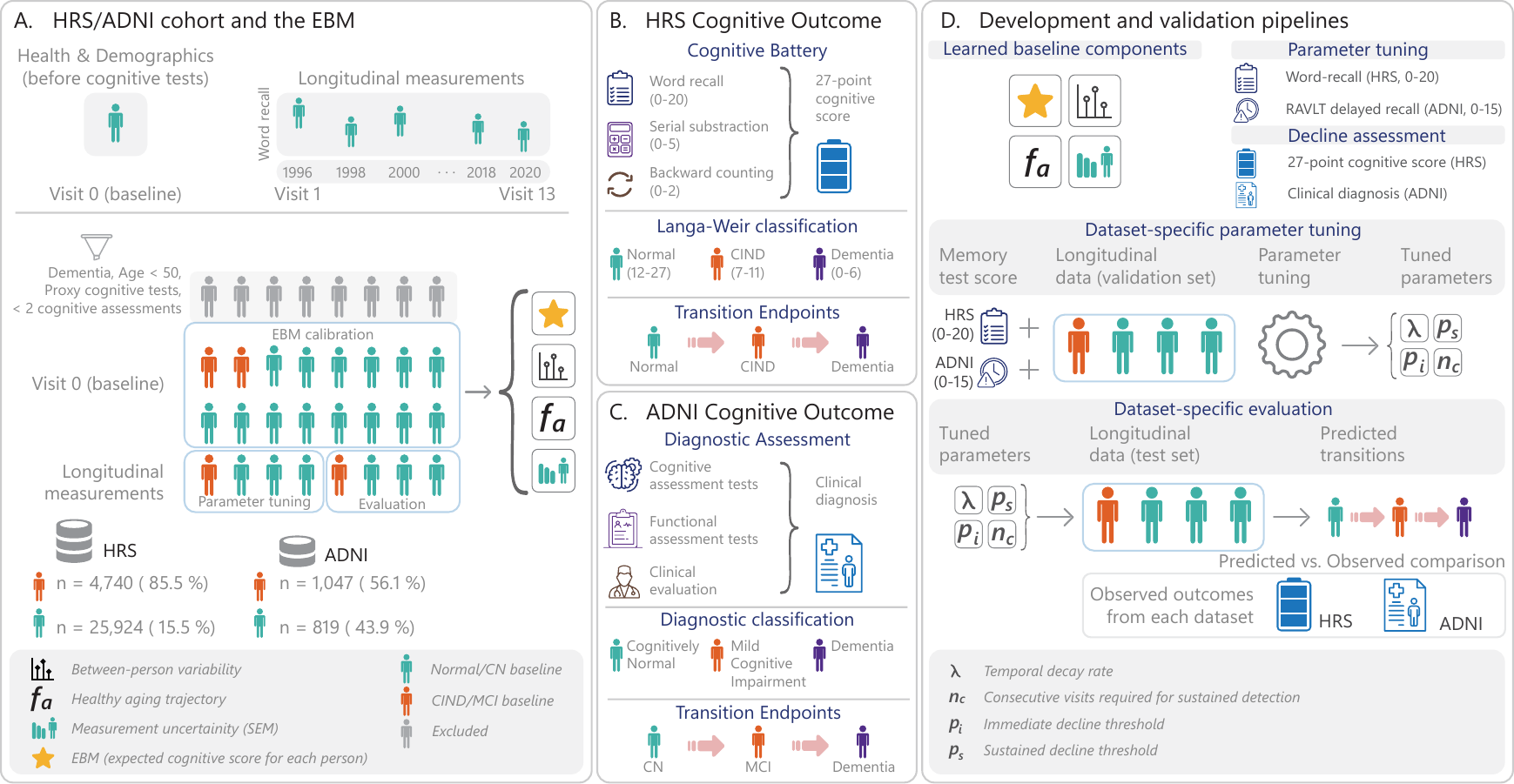}
    \caption{\textbf{Overview of the longitudinal cognitive decline detection framework across the HRS and ADNI cohorts.}
    \textbf{(A) Cohort construction, longitudinal data structure, and personalized baseline estimation.}
    Participants with baseline dementia, age $< 50$ years, proxy cognitive assessments, or fewer than two cognitive measurements were excluded. Baseline demographic and health features were used to estimate each participant's personalized EBM baseline, representing the expected memory score at study entry given their baseline profile. Repeated cognitive measurements were then used to monitor longitudinal change and detect decline. The resulting cohorts were split into parameter-tuning and evaluation subsets. During EBM training, the model learned feature-specific shape functions (see Supplementary figure~\ref{fig:shape_functions}, including the age-related function used to adjust expected performance over time; between-person variability and measurement uncertainty (SEM) were also estimated from the data.
    \textbf{(B) HRS cognitive outcome definition.}
    Word recall (0--20), serial subtraction (0--5), and backward counting were combined into the 27-point Langa--Weir total cognition score used to define Normal cognition (12--27), CIND (7--11), and Dementia (0--6) states and corresponding transition endpoints.
    \textbf{(C) ADNI cognitive outcome definition.}
    ADNI outcomes were based on clinical diagnostic classifications of cognitively normal (CN), mild cognitive impairment (MCI), and  Dementia, informed by cognitive, functional, and clinical assessments, including RAVLT delayed recall.
    \textbf{(D) Development and evaluation pipeline.}
    The same longitudinal Bayesian framework was adapted separately for HRS and ADNI using dataset-specific memory signals and outcome definitions. Parameter tuning was performed within age bands using longitudinal word-recall scores in HRS and RAVLT delayed-recall scores in ADNI. Tuned parameters included the temporal decay rate ($\lambda$), immediate decline threshold ($p_i$), sustained decline threshold ($p_s$), and the number of consecutive visits required for sustained-decline detection ($n_c$). The EBM-derived healthy-aging trajectory adjusted each participant's personalized baseline over time, while between-person variability and measurement uncertainty (SEM) governed Bayesian updating. During evaluation, tuned parameters were applied to held-out test trajectories to generate predicted cognitive transitions, which were compared with dataset-specific outcome definitions: Langa--Weir classification in HRS and clinical diagnosis in ADNI.}
    \label{fig:infographic_data}
\end{figure}

\section*{Methods}

We evaluated PRISM in two longitudinal cohorts: The Health and Retirement Study (HRS) and the Alzheimer's Disease Neuroimaging Initiative (ADNI). HRS served as the primary evaluation dataset because it provides a large, nationally representative sample of older US adults with biennial cognitive assessments collected over a 24-year period. ADNI was used as an independent external validation cohort because it includes cognitive diagnoses assigned by clinicians, allowing evaluation in a setting with an alternative outcome definition and more frequent follow-up assessments. PRISM was developed and tuned independently within each cohort using dataset-specific cognitive outcome measures and evaluation outcomes (see Figure~\ref{fig:infographic_data}; Supplementary Note~\ref{note:ground_truth}). 
\subsection*{Data}
\subsubsection*{The Health and Retirement Study (HRS)}

\paragraph{Cohort description:} Data for this cohort came from the RAND HRS Longitudinal File (2022 v1)~\cite{randhrs2025,hrs2025rand}, which is a biennial longitudinal survey examining aging and health among adults in the United States~\cite{bugliari2023randhrs}. We analyzed data from waves 3 through 15 (1996–2020), which represent the years during which all cognitive assessments needed for this study (word recall, serial subtraction, and backward counting) were administered consistently (Figure~\ref{fig:infographic_data}B). Participants for the current study were limited to individuals aged 50 years or older at baseline and were eligible if they: (1) completed a self-administered (non-proxy) cognitive interview, (2) had non-missing scores for both immediate and delayed word recall, (3) had cognitive assessments from at least two study visits, and (4) were not were not classified as having dementia at baseline according to the Langa–Weir cognitive classification algorithm. After filtering, the analytic sample included 30,664 participants contributing approximately 199,000 longitudinal observations (Figure~\ref{fig:infographic_data}A), with a mean follow-up of 6.5 visits (range 2-13). Since participants entered and exited the study at different time points, not all individuals contributed to the full 13 possible study visits across the 24-year period. Participants’ data were randomly partitioned at the individual level into non-overlapping training (EBM calibration, 70\%; n = 21,464), validation (parameter tuning, 15\%; n = 4,600), and held-out test (final evaluation, 15\%; n = 4,600) sets, such that visits from the same participant remained within the same split (Figure~\ref{fig:infographic_data}D; Supplementary Note~\ref{note:HRS}).

\paragraph{Monitored longitudinal cognitive signal} The primary longitudinal cognitive signal monitored by PRISM was the word recall composite score (0 – 20), calculated as the sum of immediate and delayed recall of a 10-word list administered at each study visit. Respondents read one of four altering word lists and asked to recall as many words as possible immediately after presentation and again after a short delay. The word lists were adapted from the Iowa Established Populations for Epidemiological Study of Elderly (EPESE)~\cite{ofstedal2005cognitive} and rotated across waves to reduce practice effects. Word recall was selected because it provided the broadest longitudinal scoring range, lowest missingness, and most consistent availability across HRS waves, while assessing episodic verbal memory, one of the earliest cognitive domains affected during age-related cognitive decline and Alzheimer's disease~\cite{elias2000preclinical,backman2005cognitive}.

\paragraph{Baseline variables:} The EBM baseline model used 24 variables measured at each participant's first study visit, spanning demographics, lifestyle, socioeconomic, medical, healthcare-utilization, sensory, mental health and functional domains. Full variable definitions and coding are provided in Supplementary Figure~\ref{fig:feature_distributions}.

\paragraph{Outcome definition:} Cognitive status at each study visit was assigned using the Langa–Weir classification~\cite{langa2020langaweir,langa2025langaweir}, which classifies participants as Normal (12–27), Cognitive Impairment No Dementia (CIND; 7–11), or Dementia (0–6) based on a 27-point total cognition composite combining word recall (0–20), serial subtraction (0–5), and backward counting (0–2) (Figure~\ref{fig:infographic_data}B). PRISM monitored longitudinal word recall trajectories to identify emerging cognitive decline, which was evaluated against subsequent worsening in Langa-Weir cognitive status.
Since the HRS Langa-Weir classification partly incorporates word recall, the monitored signal and reference endpoint are not fully independent. We therefore evaluated PRISM in ADNI using clinician-assigned diagnostic worsening based on broader clinical information and a different memory measure, providing a validation setting with reduced overlap between the monitored signal and endpoint.

\subsubsection*{The Alzheimer's Disease Neuroimaging Initiative (ADNI)}

\paragraph{Cohort description:} ADNI is a longitudinal study conducted across multiple research centres to investigate the progression of Alzheimer's disease using clinical, cognitive, imaging, genetic, and biomarker assessments~\cite{petersen2010adni,adni2024initiative}. Data were drawn from ADNI phases 1–4 (2005–2026), spanning approximately 21 years of data collection. Cognitive follow-up assessments were collected at approximately annual intervals (median interval 12 months; interquartile range 7–16 months), approximately twice as frequently as the biennial HRS assessments. Participants were included if they had valid RAVLT (Rey Auditory Verbal Learning Test)~\cite{rey1983ravlt} delayed recall scores and visit-level diagnostic classifications, contributed at least two visits with cognitive assessments, and were not classified as having dementia at baseline. After filtering, the analytic sample comprised 1,866 participants, including 819 cognitively normal (CN) and 1,047 mild cognitive impairment (MCI) at baseline, contributing 9,790 person-visit observations. Participants visited on average 5.2 times (median 5; range 2-19). Since the ADNI sample was substantially smaller than HRS, evaluation used stratified 5-fold cross-validation with individual-level splits; within each fold, participants were derived into training, validation and test sets (60\%, 20\%, and 20\%; Supplementary Note~\ref{note:ADNI}).

\paragraph{Monitored longitudinal cognitive signal:} The primary longitudinal cognitive signal monitored by PRISM was the Rey Auditory Verbal Learning Test (RAVLT) delayed recall score (0–15). The RAVLT assesses episodic verbal learning and memory using repeated learning trials, interference recall, and delayed recall after approximately 20-30 minutes.  Although structurally different from the HRS word recall test, the RAVLT delayed recall measures the same broad cognitive domain of episodic verbal memory and provides a clinically relevant longitudinal signal for monitoring cognitive decline. 

\paragraph{Baseline variables:} The EBM baseline model was restricted to eight routinely collected demographic, health, functional and mood-related variables: age, sex, years of education, race, partnership status, Functional Activities Questionnaires (FAQ) score, Geriatric Depression Scale (GDS) score, and body mass index. This restriction allowed PRISM to be evaluated using routine clinical variables without relying on specialized neuroimaging, genetic, or biomarker data.

\paragraph{Outcome definition:} For evaluation, visit-level cognitive status was defined using ADNI clinician-assigned classifications of cognitively normal (CN), mild cognitive impairment (MCI), or dementia (Figure~\ref{fig:infographic_data}C). These classifications integrate cognitive test performance, functional assessment, and clinical evaluation, rather than relying on RAVLT delayed recall score alone. PRISM monitored longitudinal RAVLT delayed recall trajectories to identify emerging cognitive decline, which was evaluated against subsequent worsening in ADNI clinical cognitive status. 

\subsubsection*{Test-retest reliability}
Both HRS word recall task and the ADNI RAVLT delayed recall measure episodic verbal memory using word-list recall tasks administered during a single testing session. To estimate within-person measurement variability for Bayesian updating, test-retest reliability was fixed at r=0.70 for both cohorts. Published studies using ADNI data report moderate-to-high test-retest reliability for RAVLT delayed recall with values ranging from 0.76 to 0.80 across 6-24 month retest intervals~\cite{hammers2022reliablechange}. For HRS, directly comparable published reliability estimates for the exact immediate and delayed word recall composite are limited, although related delayed verbal recall measures in older adults have shown reliability values within a similar range\cite{webb2022testretest}. We therefore additionally performed internal validation within the HRS training set including cognitively stable participants with at least four visits, which yielded to wave-to-wave correlation of $r=0.74$. 

The reliability value was used to calculate the standard error of measurement (SEM), according to $SEM\ =\ {SD}_y\ \sqrt{(1-r)}$, where ${SD}_y$ represents the observed standard deviation of cognitive scores. The SEM quantifies expected within-person fluctuation in observed scores due to factors such as attention, fatigue, and testing conditions, independent of genuine cognitive change.  

\subsection*{The PRISM Framework}

PRISM is a three-stage framework for monitoring individual cognitive decline that integrates interpretable machine learning with Bayesian longitudinal modeling. Stage 1 uses Explainable Boosting Machine (EBM)~\cite{nori2019interpretml}, an interpretable generalized additive model that captures transparent nonlinear effects of the features, to infer a personalized cognitive baseline from demographic, health, and functional characteristics, providing an individualized starting point when no prior of longitudinal observations are available. Stage 2 then refines this baseline dynamically by combining prior estimates with newly collected cognitive scores through Bayesian conjugate updating with an exponential time decay. This process produces a noise-reduced, recency-weighted estimate of the individual’s current cognitive status. Stage 3 estimates the likelihood of cognitive decline from the longitudinal pattern of the observed scores using Bayesian linear regression. This approach employs a prior distribution centered on zero change over time, effectively starting from the assumption of stable cognition unless the longitudinal data provide sufficient evidence to support a decline. Consequently, estimates remain cautious when only limited observations exist, but confidence in decline grows as more longitudinal measurements accumulate and offer stronger evidence of sustained change. These three stages produce three continuous, interpretable outputs: (1) a personalized expected cognitive baseline, (2) a posterior estimate of current cognitive state, and (3) a probability of cognitive decline. To translate these continuous probabilistic outputs into clinically useful decisions, we additionally specify configurable detection rules that combine the level and slope components to identify declines beyond that exceed what would be expected in healthy aging.

\begin{figure}[!ht]
    \centering
    \includegraphics[width=\textwidth]{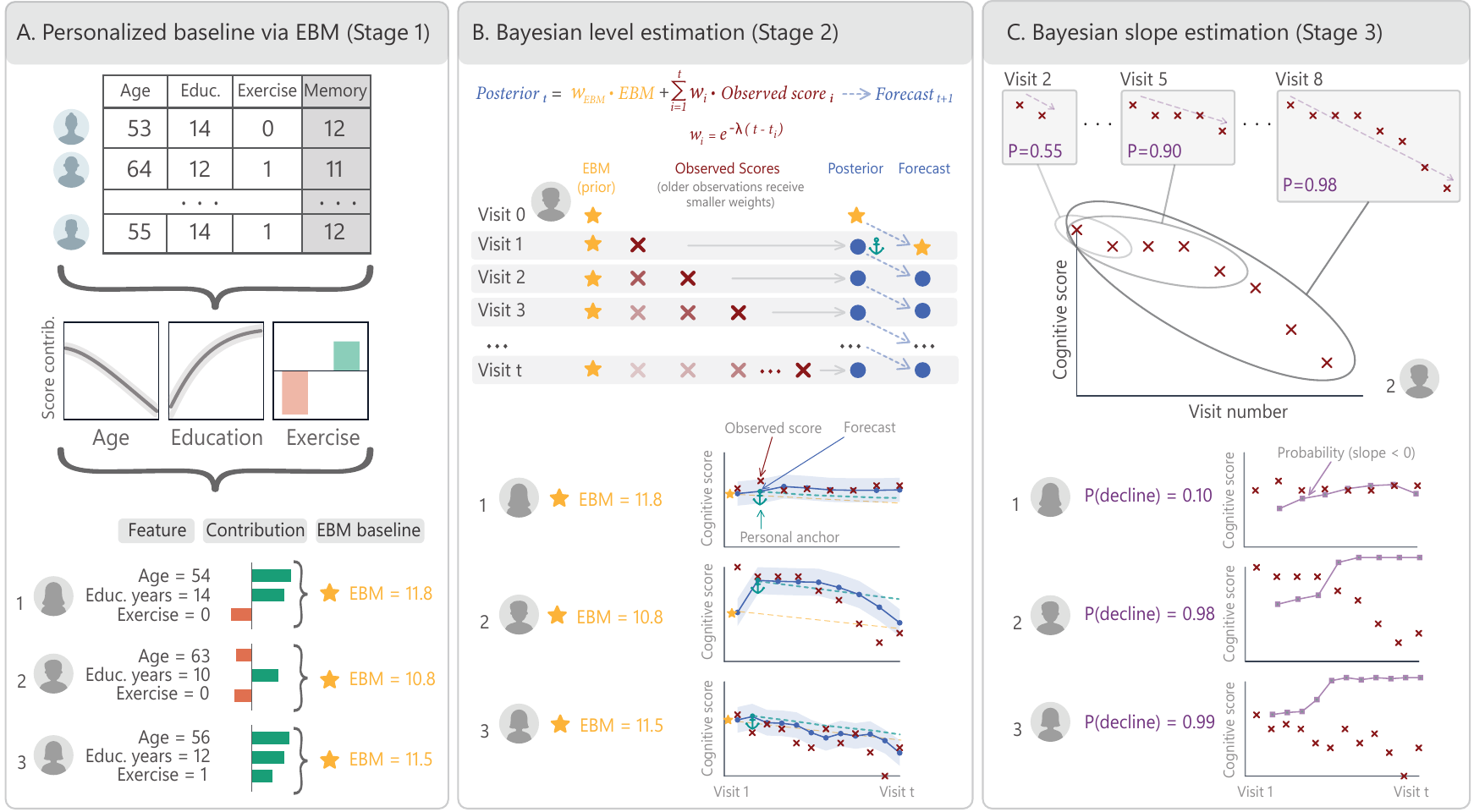}
    \caption{
\textbf{Illustrative example of the three-stage framework for personalized cognitive decline detection, shown for three individuals from the Health and Retirement Study (HRS).} 
\textbf{(A) Personalized baseline estimation via EBM.} An Explainable Boosting Machine (EBM) learns feature-specific shape functions from population-level demographic and health data and applies them to each individual to estimate an expected baseline cognitive score. Feature values contribute positively (green) or negatively (coral) to the predicted baseline, and the summed contribution plus intercept gives the EBM baseline score (yellow star). Because individuals have different feature profiles, their predicted baselines differ; for example, a 54-year-old with 14 years of education and no exercise received an EBM baseline of 11.8 on the 0--20 word recall scale.
\textbf{(B) Bayesian level estimation.} The EBM baseline serves as a population-informed prior and is combined with observed cognitive scores using precision-weighted Bayesian updating with temporal decay, so recent observations contribute more strongly than older ones. The posterior estimate (blue) is carried forward as the next-visit forecast, with uncertainty shown by the 95\% credible interval. The teal dashed line denotes the age-adjusted anchor used for below-baseline detection: at the first observed visit, the EBM-derived baseline provides the cold start reference; from the second visit onward, the reference is replaced by a personal anchor derived from the posterior after the first observed score and then adjusted over time using the EBM age function (yellow dashed line).
\textbf{(C) Bayesian slope estimation.} A Bayesian linear model estimates the posterior probability of decline, $P(\mathrm{decline})=P(\mathrm{true\ slope}<0)$. For one declining individual (participant 2), zoomed panels show the data available at visits 2, 5, and 8, with increasing confidence as observations accumulate (visit 2: $P=0.55$; visit 5: $P=0.90$; visit 8: $P=0.98$). The lower panels show $P(\mathrm{decline})$ trajectories for the three individuals: Person~1 remains stable ($P=0.10$), whereas Persons~2 and~3 show high-confidence decline ($P=0.98$ and $P=0.99$), corresponding to sharp and gradual decline patterns. $P(\mathrm{decline})$ is subsequently used by the three-tier detection rule to generate flags.
}
    \label{fig:PRISM_framework}
\end{figure}

\subsubsection*{Stage 1: Personalized baseline via Explainable Boosting Machine}
Stage 1 derives a personalized cognitive baseline using an Explainable Boosting Machine (EBM), an interpretable generalized additive model that captures nonlinear associations between baseline clinical features and cognitive outcomes, while maintaining transparency at the level of each feature (Supplementary Figure~\ref{fig:shape_functions}. Although EBMs are capable of modeling pairwise interactions between features, we disabled these interaction terms in this study to enhance interpretability and allow each feature's effect to be understood independently (Supplementary Note~\ref{note:EBM}. In contrast to black-box machine learning models, the EBM breaks down each prediction into additive effects of individual variables, allowing straightforward interpretation of how demographic, health, and functional characteristics shape the predicted cognitive score (see Supplementary Figure~\ref{fig:local_explanations}.
To construct a personalized cognitive reference point, the EBM was trained using baseline data (visit 0) to derive a target cognitive score. Visit 0 corresponds to the initial cognitive evaluation of each participant, during which demographic and health measures are collected along with the cognitive score used that serves as the prediction target; subsequent follow-up evaluations are labeled visit 1, 2, and so on. For each participant, the EBM prediction specified the personalized prior mean $(\mu_0)$, and the residual standard deviation $(\sigma_0)$ represented the portion of uncertainty not accounted for by the baseline features. This initialization addresses the cold start problem by providing a personalized expectation of cognitive performance derived from the participant's clinical profile, even before sufficient longitudinal data are available are available to determine their within-person trajectory (Figure~\ref{fig:PRISM_framework}A).

\subsubsection*{Stage 2: Estimation of the Bayesian level with temporal decay}
At each follow-up visit, the EBM-derived prior estimate was updated with newly observed cognitive test scores through Bayesian conjugate updating under a normal–normal model, which assumes that both the prior estimate and the observed cognitive scores are normally distributed and jointly inform the individual’s underlying cognitive state (Supplementary Note~\ref{note:bayesian_level_estimation}. The resulting posterior estimate represents a weighted combination of the prior expectation and the observed longitudinal data, such that the information with lower uncertainty has a greater influence on the final value (Figure~\ref{fig:PRISM_framework}B). As more observations accumulated, the estimate progressively shifted away from reliance on the population-derived prior baseline and increasingly aligned with the observed longitudinal pattern of the individual participant. 
To account for evolving cognitive status over time, earlier observations are weighted using an exponential temporal decay. Specifically, an observation recorded $\Delta t$  years before the current visit receives weight $e^{-\lambda\Delta t}$ where $\lambda$ controls at which rate older observations lose influence over time; higher values produce faster down-weighting of past scores
(Supplementary Note~\ref{note:lambda}. This approach allows recent assessments to carry greater influence than those in the past, while naturally accommodating irregular follow-up intervals without assuming fixed visit spacing. The decay parameter $\lambda$ was selected on the validation set by minimizing one-step-ahead forecasting error, using grid search 
Measurement uncertainty was estimated using the standard error of measurement (SEM), derived from the test--retest reliability of the cognitive outcome ($r$) and the observed score variance ($\sigma_0^2$). The corresponding measurement precision, $\tau_m = 1/\mathrm{SEM}^2$, and prior precision, $\tau_0 = 1/\sigma_0^2$, jointly determined the relative influence of the observed longitudinal scores and the EBM-derived prior. Importantly, more reliable cognitive tests produce lower measurement uncertainty, allowing the framework to rely more strongly on observed individual trajectories, whereas noisier observations lead to greater reliance on the prior until sufficient longitudinal evidence accumulates (Figure~1C). This probabilistic framework enabled a gradual transition from population-informed initialization to increasingly personalized estimation as longitudinal observations accumulated.

\subsubsection*{Stage 3: Bayesian slope estimation}
To determine whether an individual’s observed cognitive trajectory shows evidence of decline, we estimated the slope of cognitive scores over time using ordinary least squares and apply Bayesian shrinkage to stabilize estimates when longitudinal follow-up is limited. In practice, this implies that when only a few follow-up visits are available, the model assumes cognitive stability unless there is clear evidence of decline. As additional visits accumulate, the estimated trajectory becomes increasingly driven by the individual's observed pattern of change. The prior distribution is centered on no change over time, corresponding to an initial assumption of stable cognition (Supplementary Note~\ref{note:slope_estimation}. As a result, when only two or three observations are available, the estimated probability of decline remains close to 0.5, indicating true uncertainty rather than an early, potentially misleading sign of decline (Figure~\ref{fig:PRISM_framework}C). As additional visits accumulate, the observed longitudinal trajectory progressively dominates the prior, allowing the estimated probability of decline to converge toward 0 (stable trajectory) or 1 (declining trajectory).  

Since participants age during longitudinal follow-up, the expected cognitive baseline was re-evaluated at each visit based on their age at that time. This helps to ensure that healthy age-related changes in cognition are not misinterpreted as pathological decline. Rather than relying on external normative tables, the adjustment is based directly on the nonlinear age effect learned by the EBM, enabling the expected level of cognitive performance to update dynamically as individuals grow older. The decline was therefore evaluated in relation to each participant's age-adjusted trajectory, ensuring that the gradual changes expected in healthy aging were not classified as pathological decline. 

\subsection*{Framework outputs and evaluation}
The three stages provide a probabilistic framework for estimating personalized cognitive trajectories and assessing the likelihood of cognitive decline over time. At each visit, the model produces three continuous outputs for each individual: (1) an age-adjusted expected cognitive score derived from the EBM, (2) a Bayesian posterior estimate of the current cognitive stage along with its uncertainty, and (3) the posterior probability of decline, P(decline).

To contextualize PRISM's  forecasting performance, we compared its one-step-ahead predictions against two reference approaches representing different levels of personalization: a demographic-adjusted norm baseline and a cumulative individual average (Figure 1C; Supplementary Note~\ref{note:evaluation_metrics}). The demographic-adjusted norm baseline used  cognitive scores stratified by age band (5-year intervals), education level (5 categories) and sex, following the demographic scheme proposed by Hurd et al~\cite{hurd2013monetary}. Each participant’s expected score was updated at every visit according to their current age band, allowing the baseline to account for normative cognitive aging over time while remaining population-based and non-personalized. The cumulative average baseline used the arithmetic mean of each participant's previously observed cognitive scores, providing a simple personalized forecasting approach without probabilistic updating, uncertainty modeling, or temporal weighting.

Since longitudinal cognitive monitoring involves more than minimizing next score prediction error alone, we additionally evaluated how well the framework could differentiate, at a given visit, between individuals whose cognitive status had worsened relative to their own baseline and those who had remained stable at a given visit. This was assessed using the per-visit area under the receiver operating characteristic curve (AUC). For this analysis, PRISM was compared with the standard linear mixed-effects (LME) model, as an established repeated-measures approach for longitudinal trajectory estimation. The LME model used participant-specific random intercepts and random slopes for time, with fixed effects for years since baseline assessment, baseline age, sex, and years of education (Figure 1D; Supplementary Note~\ref{note:evaluation_metrics}). For PRISM, discrimination scores were based on the posterior probability of decline, P(decline). For LME, discrimination scores were derived from participant-specific negative longitudinal slope estimates such that steeper negative slopes corresponded to stronger evidence of decline. The demographic-adjusted norm and cumulative average baselines were included as forecasting references only because they do not provide a statistical framework to estimate individual decline trajectories or uncertainty around longitudinal change. 

\subsubsection*{Three-tier detection rule}
To translate these continuous probabilistic outputs into actionable clinical alerts, a separate set of detection rules can be applied (Supplementary Note~\ref{note:three_tier_rule}. This decision layer is separable from the core Bayesian framework and can be adapted to different clinical settings, populations, dementia subtypes, or monitoring objectives. Different applications may require different thresholds, flagging strategies, or definitions of clinically meaningful decline.

The three-tier rule described below is one such implementation, designed to capture distinct temporal patterns of decline, including abrupt deterioration, immediate decline, and gradual sustained decline. Detection thresholds were optimized to prioritize early detection of sustained decline while maintaining a minimum specificity constraint among individuals who remained cognitively stable throughout follow-up. Importantly, metrics such as false-positive and false-negative rates are attributes of this adjustable decision layer, rather than of the underlying probabilistic model itself. Consequently, the detection rule can be modified, expanded with additional decline profiles, or recalibrated for new populations without altering the Bayesian estimation components.
The detection rule integrates both trajectory and level signals to identify decline beyond that expected from healthy aging. Detection requires two joint conditions: (1) a slope condition, in which $P(decline)$ exceeds a predefined threshold, indicating evidence of a downward trajectory, and (2) a level condition, in which current cognitive performance falls below a personalized age-adjusted reference level by at least a specified tolerance margin (Supplementary Note~\ref{note:det_rule_optimization}. 
The age-adjusted reference level was estimated using a hybrid approach. Initially, the reference level was based on the EBM-derived baseline, providing a cold start estimate before individual longitudinal observations were available. After the first observed memory score was incorporated, PRISM transitioned to a personalized anchor defined by the first Bayesian posterior estimate. This anchor was then adjusted over time using the age-related function learned by the EBM during training, allowing the framework to distinguish expected aging-related change from potential pathological decline (Supplementary Note~\ref{note:age_adjusted_anchor}. Requiring both conditions prevents the decline from being identified solely based on a downward trend when performance remains near the individualized age-adjusted reference level, while also reducing false alarms driven by isolated low scores without evidence of sustained decline. The level condition was evaluated using both the Bayesian posterior estimate and the current observed cognitive score, since posterior updating incorporates prior information and may respond gradually to abrupt cognitive decline.

\paragraph{Tier 0 (Safety net)} If the posterior estimate drops more than two standard deviations below the age-adjusted personal baseline and $P(decline)\geq0.50$, the individual is flagged immediately. Both criteria are fixed rather than tuned empirically: a shift of two-standard-deviations represents a statistically clear departure from the expected cognitive range, while requiring $P(decline)\geq0.50$ ensures at least moderate evidence of a downward trajectory. This tier is intended to identify sudden, substantial drops in cognitive level, such as a marked worsening between two visits, where the magnitude of the level change alone justifies immediate attention, even if the slope estimate has not yet accumulated sufficient longitudinal evidence to be considered highly reliable.

\paragraph{Tier 1 (Immediate)} When P(decline) is above a high probability threshold, and the individual’s current score is at least the specified margin below their age-adjusted personal baseline, a decline is marked for that visit. Unlike Tier 0, which reacts to extreme level deviations in the level but relies on only moderate evidence from the slope, Tier 1 requires strong trajectory evidence (a high $P(decline)$) combined with a less extreme deviation in the level. This tier includes situations where several observations together demonstrate a clear downward trend, even if no individual score indicates a dramatic drop. One visit that satisfies both criteria is enough for detection.

\paragraph{Tier 2 (Sustained)} If $P(decline)$ exceeds a lower probability threshold and the individual’s score stays below baseline for a specified number of consecutive visits, a decline is indicated. Since this threshold is set lower than in Tier 1, a single visit could simply reflect measurement noise or a short-term fluctuation rather than a genuine decline. By requiring the pattern to persist across several consecutive visits, transient changes are filtered out while maintaining sensitivity to gradual, sustained decline.

\paragraph{Recovery and unflagging:} The original purpose of the recovery and unflagging rules was to enable the framework to differentiate lasting deterioration from short-term variations. Under this approach, individuals stayed flagged until later visits no longer provided sufficient evidence of decline at any detection tier. Flags arising from sustained evidence of decline (Tier 2) were only removed after two consecutive clear visits required, while flags raised by evidence from a single visit (Tiers 0 and 1) were removed after just one clear visit. This asymmetry was intentional and embodies the idea that stronger evidence of decline should be matched by stronger evidence of recovery before a flag is lifted.
In the HRS cohort, in which cognitive status was determined using the Langa-Weir classification and short-term variability in scores was more frequent, this recovery mechanism helped to ensure that isolated low scores were not misinterpreted as sustained decline. In contrast, in ADNI, cognitive status was based on longitudinal clinical diagnosis, which showed substantially lower rates of diagnostic reversion. As a result, the unflagging option was disabled in the ADNI implementation.  

\paragraph{Parameter Optimization}

To distinguish sustained cognitive decline from temporary fluctuations in diagnostic status, the optimization was based on a longitudinal sustained-decline endpoint rather than isolated worsening events. Participants were considered endpoint-positive if they met either of the following conditions: (1) showed a decline in cognitive status compared with baseline at two or more follow-up visits, whether or not those visits were consecutive, or (2) continued to perform worse than baseline at their last recorded visit. The latter criterion was included to accommodate right-censored trajectories in which later recovery could not be observed. Participants who experienced only a single, isolated worsening visit and then returned to their baseline cognitive status were not classified as sustained decliners during optimization. This approach was intended to avoid excessive penalization of transient diagnostic fluctuations while maintaining the ability to detect genuine, persistent decline.

Optimization was carried out hierarchically in several stages (Supplementary Note~\ref{note:det_rule_optimization}. In the first stage, the temporal decay parameter $(\lambda)$, which controls the influence of older observations on Bayesian updating, was optimized globally using one-step-ahead forecasting errors on validation set. A range of candidate values was evaluated based on the mean absolute error (MAE) of longitudinal prediction, and the value that yielded the lowest MAE validation was chosen. 
Subsequently the thresholds for detecting decline were refined using a hierarchical optimization procedure aimed at maximizing early detection while keeping false alarms to a minimum. Candidate parameter configurations were initially required to meet a minimum specificity threshold among participants who maintained clinical stability over the entire follow-up period. Across all eligible configurations, optimization prioritized earlier identification of future decline using an exponentially saturating lead-time reward, that favored earlier detection without creating an overly strong bias toward extremely early alerts. Secondary selection prioritized higher $F_2$ scores to place greater weight on sensitivity. We also imposed constraints so that immediate-decline required higher probability thresholds than those indicating a sustained decline $(p_{immediate}>p_{sustained})$, thereby maintaining the intended hierarchy of alert severity. 
Finally, we conducted age-stratified analyses within four predefined age categories (50--64, 65--74, 75--84, and $\geq 85$ years) to account for variation in expected decline patterns across later life (Supplementary Note~\ref{note:age_stratified}.

\section*{Acknowledgement}
We are grateful to Dr. Catherine Diaz-Asper for her generous time, collegial support, and helpful discussions during the development of this work.

\section*{Contributions}
B.E. provided the conceptual foundation and clinical framing for integrating population-informed baselines with individualized longitudinal monitoring. M.S. developed the PRISM framework, formalized the methodological approach, conducted the analyses, prepared the figures, and drafted the manuscript. M.S. and B.E. jointly refined the methodology, interpreted the findings, revised the manuscript, and approved the final version.
\section*{Data Availability}

Processed and anonymized data and processing scripts generated for this study are available from the corresponding author upon reasonable request.

The data that support the findings of this study were obtained from the Health and Retirement Study (HRS) and the Alzheimer’s Disease Neuroimaging Initiative (ADNI). HRS data are publicly available from the University of Michigan via the HRS repository (\url{https://hrsdata.isr.umich.edu/}). The HRS is sponsored by the National Institute on Aging (grant number NIA U01AG009740) and is conducted by the University of Michigan. The RAND HRS Longitudinal File 2022 (V1) was produced by the RAND Center for the Study of Aging, with funding from the National Institute on Aging and the Social Security Administration.

The external validation data from the Alzheimer’s Disease Neuroimaging Initiative (ADNI) are available from the Alzheimer’s Disease Neuroimaging Initiative repository (\url{https://adni.loni.usc.edu/}), subject to data access approval and the ADNI data use agreement. Data used in preparation of this article were obtained from the ADNI database. ADNI was launched in 2003 as a public--private partnership, led by Principal Investigator Michael W. Weiner, MD, to support research on the progression of mild cognitive impairment and early Alzheimer’s disease through imaging, biomarker, clinical, and neuropsychological data. ADNI investigators contributed to the design and implementation of ADNI and/or provided data, but did not participate in the analysis or writing of this report. A complete listing of ADNI investigators can be found in the \href{http://adni.loni.usc.edu/wp-content/uploads/how_to_apply/ADNI_Acknowledgement_List.pdf}{ADNI Acknowledgement List}.

Data collection and sharing for ADNI is funded by the National Institute on Aging (National Institutes of Health Grant U19AG024904). The grantee organization is the Northern California Institute for Research and Education. ADNI has also received funding from the National Institute of Biomedical Imaging and Bioengineering, the Canadian Institutes of Health Research, and private-sector contributions through the Foundation for the National Institutes of Health.


\section*{Code Availability}

The code developed to implement and evaluate the PRISM framework is publicly available at \url{https://github.com/MariaSahakyan/PRISM}. The repository includes the code required to reproduce the principal analyses and evaluations reported in this study.

\pagebreak
\bibliography{bibliography}
\bibliographystyle{unsrtnat}
\pagebreak


\beginsupplement

\begin{center}
{\Large\bfseries Supplementary Materials}\\
\end{center}


\input{arxiv_supplementary}

\clearpage

\end{document}

%% file: arxiv_supplementary.tex
\section*{Supplementary Figures}

\begin{figure}[!ht]
    \centering
    \includegraphics[width=\textwidth]{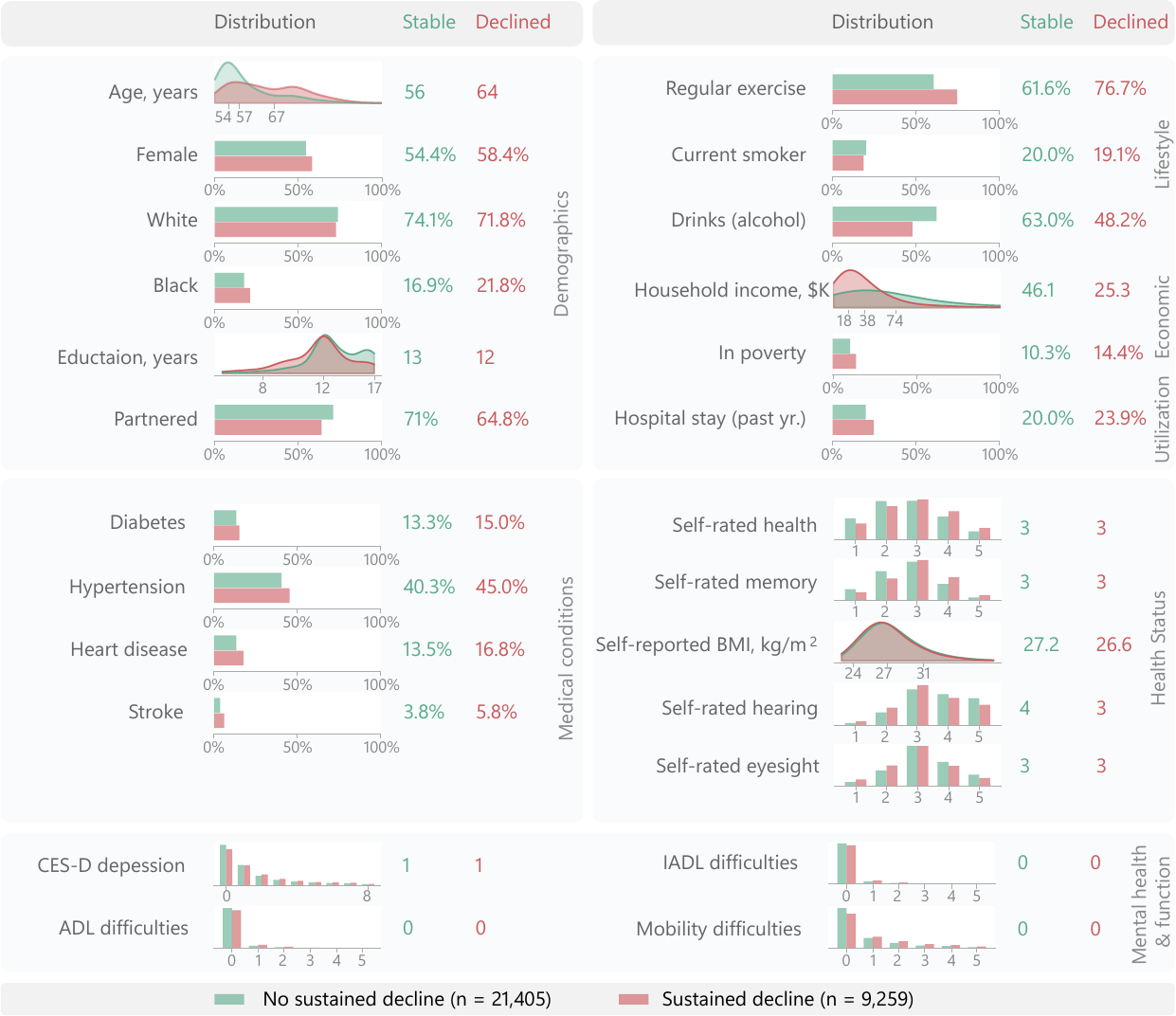}
    \caption{\textbf{Distribution of baseline input features across participants with and without sustained cognitive decline (HRS dataset).}
    The figure summarizes demographic, lifestyle, socioeconomic, medical, functional, and self-reported health variables used as model inputs. Green represents participants without sustained decline (n = 21,405), while red represents participants with sustained decline (n = 9,259). Continuous variables are shown as density distributions, binary variables as percentages, and ordinal/count variables as histograms. The values displayed on the right correspond to group summary statistics: means for continuous variables, medians for ordinal variables, and percentages for binary variables. For continuous-feature panels, x-axis ticks indicate the 5th percentile, median, and 95th percentile of the pooled distribution across all participants. Although participants with sustained decline differed on average across multiple domains, including age, education, income, medical burden, self-rated health, and functional status, the distributions also show substantial overlap between groups. These descriptive patterns highlight the heterogeneous baseline profiles from which cognitive trajectories emerge and provide context for estimating expected cognition at the individual level before evaluating longitudinal change. All features are extracted from the RAND HRS Longitudinal dataset
    }
    \label{fig:feature_distributions}
\end{figure}

\begin{figure}[!ht]
    \centering
    \includegraphics[width=\textwidth]{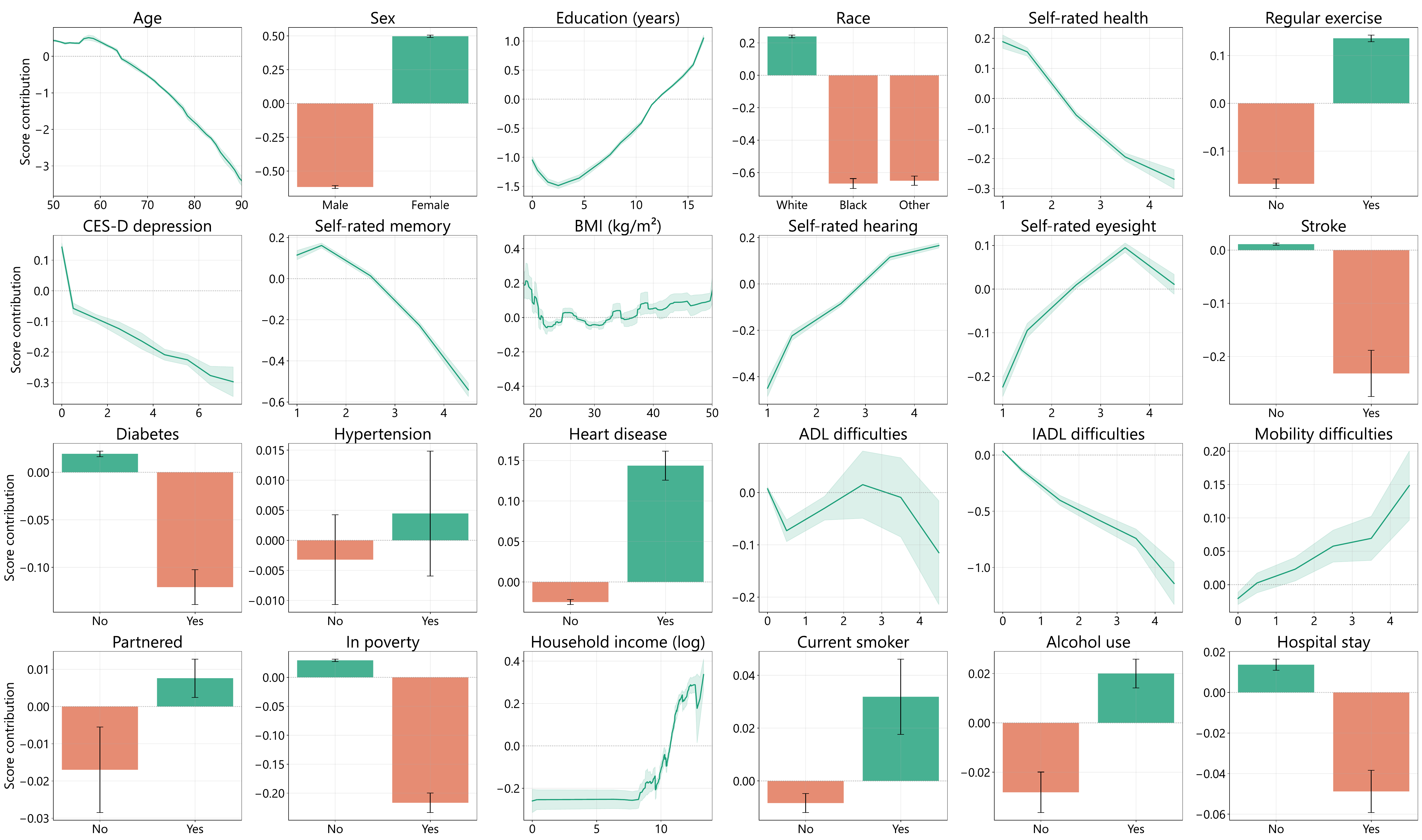}
\caption{
\textbf{EBM shape functions for the 24 predictors of baseline word recall score in HRS.} Each panel shows the learned univariate contribution of one feature to the predicted word recall score on the 0--20 scale. Positive values increase the predicted baseline score, whereas negative values decrease it. Lines show continuous feature functions and bars show categorical or binary feature contributions; shaded bands and error bars indicate 95\% confidence intervals derived from 16 outer bagging iterations. The learned functions show expected major patterns, including lower predicted scores with older age, higher predicted scores with more education, and lower predicted scores with higher depressive symptoms, poorer self-rated memory, stroke, diabetes, poverty, and recent hospital stay. Household income is shown on the transformed log scale. Wider intervals for sparse regions, including higher functional-difficulty counts and extreme BMI values, indicate greater uncertainty. Shape functions represent predictive associations learned from the training data rather than causal effects; small or counterintuitive feature contributions may reflect covariate correlations, residual confounding, or limited independent signal after adjustment by the additive model.
}
    \label{fig:shape_functions}
\end{figure}

\begin{figure}[!ht]
    \centering
    \includegraphics[width=\textwidth]{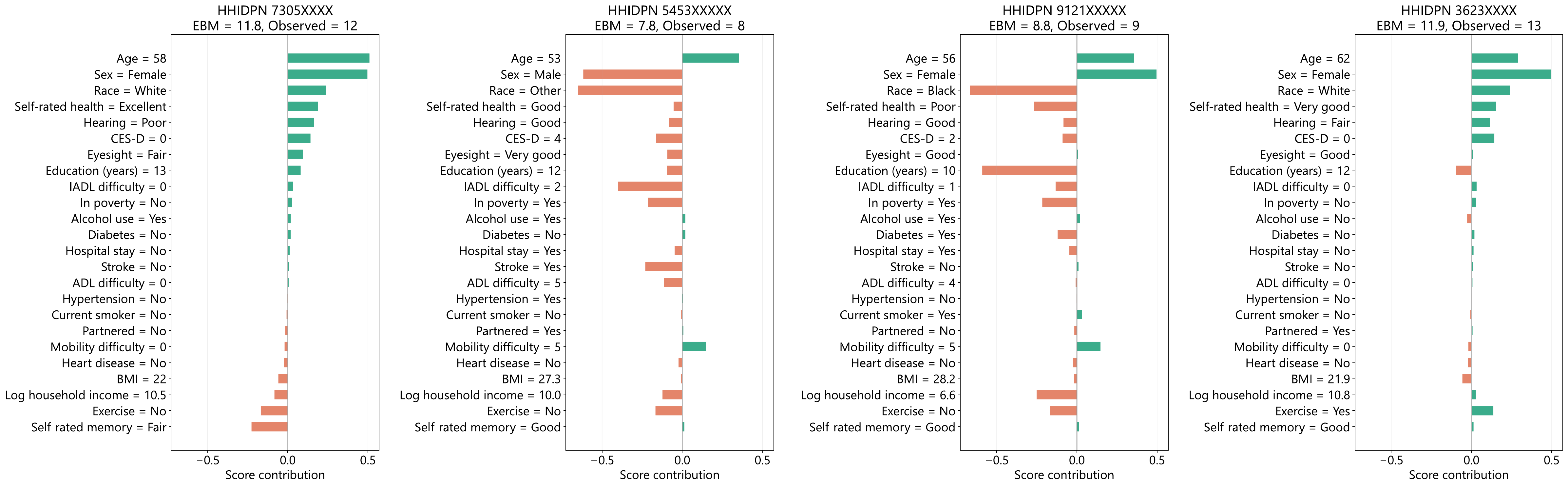}
    \caption{
\textbf{Local EBM explanations for four representative participants.} Each panel decomposes the EBM-predicted baseline word recall score at the first visit into feature-level contributions. Green bars indicate positive contributions to the predicted score, whereas coral bars indicate negative contributions. Features are shown in a common order across panels, ranked by contribution magnitude for the first participant, to facilitate visual comparison. The full EBM prediction equals the model intercept plus the sum of all feature contributions. These personalized decompositions illustrate how the model forms participant-specific baseline expectations before longitudinal Bayesian monitoring.
}
    \label{fig:local_explanations}
\end{figure}

\clearpage

\section*{Supplementary Methods}

\section{The PRISM framework}\label{note:PRISM}
\subsection{Personalized Baseline via Explainable Boosting Machine}\label{note:EBM}
The personalized baseline is estimated using an Explainable Boosting Machine (EBM), a cyclic gradient-boosted generalized additive model that learns feature-specific shape functions. In its more general GA\textsuperscript{2}M form, an EBM can include both univariate feature terms and selected pairwise interaction terms:

\begin{equation}
\hat{y}_i \;=\; \beta_0 + \sum_{j=1}^{p} f_j(x_{ij}) 
+ \sum_{j<k} f_{jk}(x_{ij}, x_{ik}).
\label{eq:ebm_general}
\end{equation}

\noindent Here, $i$ indexes individuals, $j$ and $k$ index input features, $\hat{y}_i$ denotes the predicted cognitive score for individual $i$, $x_{ij}$ denotes the value of feature $j$ for individual $i$, $f_j(\cdot)$ is the learned univariate shape function for feature $j$, and $f_{jk}(\cdot,\cdot)$ denotes a potential pairwise interaction function between features $j$ and $k$.

\noindent In our implementation, pairwise interactions were disabled 
(\texttt{interactions\,=\,0}), reducing the model to the purely additive GAM form:

\begin{equation}
\hat{y}_i \;=\; \beta_0 + \sum_{j=1}^{p} f_j(x_{ij}).
\label{eq:ebm}
\end{equation}

\noindent This additive specification preserves interpretability at both the global and local levels: each shape function $f_j(\cdot)$ describes the contribution of a feature across its observed range, while each individual prediction can be decomposed into the intercept and participant-specific feature contributions.

\noindent The EBM was trained on baseline data, defined as visit 0, using the observed cognitive score as the prediction target. Visit 0 corresponds to the initial cognitive assessment of each participant, when demographic and health measures were collected along with the cognitive outcome; subsequent follow-up assessments were labeled visit 1, 2, and so on. The resulting prediction of EBM, denoted $\hat{y}_i$, defines the participant-specific prior mean $\mu_{0,i}$ used in the Bayesian updating step:

\begin{equation}
\mu_{0,i} = \hat{y}_i.
\label{eq:mu0}
\end{equation}
\noindent This baseline model provides a personalized expected cognitive score before longitudinal observations are available, which addresses the cold-start problem. Prior uncertainty was estimated from the in-sample EBM residuals on the training set, comparing each participant's observed baseline cognitive score with the corresponding EBM-predicted baseline score:

\begin{equation}
\sigma_0 = \mathrm{SD}\!\left(y_i - \hat{y}_i\right),
\label{eq:sigma0}
\end{equation}

\noindent where $y_i$ denotes the observed cognitive score at baseline visit 0 and $\hat{y}_i$ denotes the EBM-predicted baseline score. This residual standard deviation captures baseline variability not explained by the available EBM features; its square defines the prior variance in the Bayesian updating step.
\paragraph{EBM hyperparameters.}
The same EBM configuration was used in both cohorts:\\ \texttt{max\_bins\,=\,256}, 
\texttt{outer\_bags\,=\,16}, 
\texttt{inner\_bags\,=\,0}, 
\texttt{max\_rounds\,=\,1500},\\ \texttt{learning\_rate\,=\,0.02}, 
\texttt{interactions\,=\,0}, and \texttt{random\_state\,=\,42}.
\subsection{Bayesian Level Estimation with Temporal Decay}\label{note:bayesian_level_estimation}

We model each individual's underlying cognitive ability $\theta_i$ as a latent trait inferred from repeated cognitive test scores using Normal--Normal conjugate Bayesian updating.

\paragraph{Prior and likelihood.}
The EBM provides a point estimate of each participant's expected baseline cognitive score. This estimate was used as the center of the participant-specific prior distribution, such that $\mu_{0,i}=\hat{y}_i$. The prior and observation model are given by:

\begin{align}
\theta_i &\sim \mathcal{N}\!\left(\mu_{0,i},\;\sigma_0^2\right) \label{eq:prior}\\[4pt]
y_{it} \mid \theta_i &\sim \mathcal{N}\!\left(\theta_i,\;\mathrm{SEM}^2\right), \label{eq:likelihood}
\end{align}

\noindent where $\theta_i$ denotes the latent cognitive level for individual $i$, $\mu_{0,i}$ is the EBM-derived prior mean, $\sigma_0$ is the EBM residual standard deviation (Eq.~\ref{eq:sigma0}), $y_{it}$ is the observed cognitive score for individual $i$ at visit~$t$, and $\mathrm{SEM}$ is the standard error of measurement:

\begin{equation}
\mathrm{SEM} = \mathrm{SD}_y \cdot \sqrt{1 - r}.
\label{eq:sem}
\end{equation}

\noindent Here, $\mathrm{SD}_y$ is the standard deviation of the target cognitive score in the training set and $r$ is the test--retest reliability coefficient. The two variance terms capture distinct sources of uncertainty: $\sigma_0^2$ represents baseline prediction uncertainty around the EBM-derived prior mean, whereas $\mathrm{SEM}^2$ represents within-person measurement-error variance.

\paragraph{Precision-weighted update with temporal decay.}

We define the prior precision as $\tau_0 = 1/\sigma_0^2$ and the measurement precision as $\tau_m = 1/\mathrm{SEM}^2$. Let $t$ denote the current visit index $(t = 1,2,\ldots)$, and let $d_{ij}$ denote the calendar date of visit~$j$ for individual~$i$. Differences between visit dates were converted to years by dividing the elapsed number of days by 365.25. To capture cognitive change over time, older observations were exponentially down-weighted. When computing the posterior at visit~$t$, each observation $y_{ij}$ receives weight

\begin{equation}
w_{ij}^{(t)} = \exp\!\left[-\lambda \left(d_{it} - d_{ij}\right)\right],
\label{eq:decay}
\end{equation}

\noindent where $d_{it}-d_{ij}$ is the elapsed time in years between visit~$j$ and the current visit~$t$, and $\lambda > 0$ is the temporal decay rate. The current observation has $d_{it}-d_{it}=0$ and therefore receives weight 1, while earlier observations receive progressively smaller weights.

\noindent Since the prior and weighted likelihood contributions are Normal, the updated posterior distribution remains Normal and can be computed analytically. At visit~$t$, the posterior precision $\tau_{\mathrm{post},it}$, posterior mean $\mu_{\mathrm{post},it}$, and posterior standard deviation $\sigma_{\mathrm{post},it}$ for individual~$i$ are:

\begin{align}
\tau_{\mathrm{post},it} &= \tau_0 + \sum_{j=1}^{t} w_{ij}^{(t)}\,\tau_m \label{eq:tau_post}\\[4pt]
\mu_{\mathrm{post},it} &= 
\frac{\tau_0\,\mu_{0,i} \;+\; \displaystyle\sum_{j=1}^{t} w_{ij}^{(t)}\,\tau_m\,y_{ij}}
{\tau_{\mathrm{post},it}} \label{eq:mu_post}\\[4pt]
\sigma_{\mathrm{post},it} &= \frac{1}{\sqrt{\tau_{\mathrm{post},it}}}.
\label{eq:sigma_post}
\end{align}

\subsection{Bayesian Slope Estimation}\label{note:slope_estimation}

We estimated the rate of cognitive change over time and combined the observed slope with a zero-centered prior to stabilize slope estimates when follow-up data were limited. The EBM-derived prior served as the initial personalized baseline before sufficient longitudinal observations were available. After the first observed cognitive score was incorporated, it established the individual's personal anchor; however, slope estimation was performed only from the next eligible visit onward, when at least two observed cognitive scores were available.

\noindent At each eligible visit~$t$, we computed the ordinary least-squares (OLS) slope using all observed cognitive scores available up to that visit. Let $d_{i1}$ denote the date of the anchor visit, defined as the first observed cognitive assessment after EBM prior initialization, and let $a_{ij}=d_{ij}-d_{i1}$ denote elapsed time in years since that anchor visit. The OLS slope was computed as:

\begin{align}
\hat{\beta}_{it} &= 
\frac{\displaystyle\sum_{j=1}^{t} (a_{ij} - \bar{a}_{it})(y_{ij} - \bar{y}_{it})}
{\displaystyle S_{aa,it}}, 
\qquad 
S_{aa,it} = \sum_{j=1}^{t} (a_{ij} - \bar{a}_{it})^2,
\label{eq:beta_hat}\\[4pt]
\mathrm{se}_{\beta,it} &= \frac{\mathrm{SEM}}{\sqrt{S_{aa,it}}}.
\label{eq:se_beta}
\end{align}

\noindent Here, $\bar{a}_{it}$ and $\bar{y}_{it}$ are the within-person means of elapsed time and observed cognitive score over visits $1,\ldots,t$, and $\hat{\beta}_{it}$ is the estimated annual rate of cognitive change in points per year. The slope standard error $\mathrm{se}_{\beta,it}$ uses the standard error of measurement, $\mathrm{SEM}$, as the observation-level noise estimate.

\noindent We placed a zero-centered Normal prior on the true individual slope, representing an initial expectation of no systematic change before sufficient longitudinal evidence is available:

\begin{equation}
\beta_i \sim \mathcal{N}\!\left(0,\;\sigma_\beta^2\right),
\label{eq:slope_prior}
\end{equation}

\noindent where $\sigma_\beta$ is the empirical standard deviation of within-person OLS slopes estimated from training participants with at least three visits. This prior shrinks unstable early slope estimates toward no change while allowing stronger evidence of decline or improvement as longitudinal data accumulate.

\noindent Combining this prior with the OLS slope estimate yields a Normal posterior distribution for the slope. Defining the prior and data precisions as

\begin{equation}
\tau_{\beta,\mathrm{prior}} = \frac{1}{\sigma_\beta^2},
\qquad
\tau_{\beta,\mathrm{data},it} = \frac{1}{\mathrm{se}_{\beta,it}^2},
\label{eq:slope_precisions}
\end{equation}

\noindent the posterior precision, mean, and standard deviation at visit~$t$ are:

\begin{align}
\tau_{\beta,\mathrm{post},it} &= 
\tau_{\beta,\mathrm{prior}} + \tau_{\beta,\mathrm{data},it}
\label{eq:tau_beta}\\[4pt]
\beta_{\mathrm{post},it} &= 
\frac{\tau_{\beta,\mathrm{data},it}\,\hat{\beta}_{it}}
{\tau_{\beta,\mathrm{post},it}}
\label{eq:beta_post}\\[4pt]
\sigma_{\beta,\mathrm{post},it} &= 
\frac{1}{\sqrt{\tau_{\beta,\mathrm{post},it}}}.
\label{eq:sigma_beta_post}
\end{align}

\noindent The posterior probability of decline was computed as the probability that the slope is below zero:

\begin{equation}
P(\mathrm{decline}_{it}) = 
\Phi\!\left(\frac{-\beta_{\mathrm{post},it}}
{\sigma_{\beta,\mathrm{post},it}}\right),
\label{eq:p_decline}
\end{equation}

\noindent where $\Phi$ denotes the standard Normal cumulative distribution function. With few observations, the posterior slope is shrunk toward zero, keeping $P(\mathrm{decline}_{it})$ close to 0.5 and reflecting uncertainty. As follow-up observations accumulate, data precision increases and the posterior slope increasingly reflects the observed trajectory.

\section{Early detection of cognitive decline}\label{note:detection}
\subsection{Age-Adjusted Baseline and Personal Anchor}\label{note:age_adjusted_anchor}

Cognitive test performance is expected to change with age. To avoid flagging age-expected decline as abnormal decline, we adjusted the EBM-derived baseline at each visit using the age-specific shape function learned by the EBM. Let $\alpha_i$ denote individual~$i$'s age at baseline and let $\Delta t$ denote elapsed time in years since baseline. The age-adjusted EBM baseline at time~$t$ was defined as:

\begin{equation}
\mu_{0,i}^{(\mathrm{adj})}(t) = \mu_{0,i} + 
\bigl[f_{\mathrm{age}}(\alpha_i + \Delta t) - f_{\mathrm{age}}(\alpha_i)\bigr],
\label{eq:age_adj}
\end{equation}

\noindent where $\mu_{0,i}$ is the EBM-derived prior mean and $f_{\mathrm{age}}(\cdot)$ is the EBM shape function for age. This adjustment shifts the expected baseline according to the model's learned age--cognition relationship. If the learned age function predicts lower scores at older ages, the reference baseline declines accordingly, so that decline is evaluated relative to age-expected change rather than against a fixed baseline. No external normative table was used; the adjustment was derived entirely from the trained EBM.

\paragraph{Personal anchor.}
Before any observed cognitive score is available, the EBM-derived baseline serves as the initial personalized reference. At visit~1, the first observed cognitive score is incorporated through the Bayesian update, producing the posterior mean $\mu_{\mathrm{post},i1}$. This posterior mean defines the individual's personal anchor. The anchor is then used as the comparison reference from visit~2 onward, when longitudinal change can be evaluated relative to the individual's own observed starting level.

\noindent To keep the anchor comparable with later visits, it is shifted by the age-related change learned by the EBM. Let $a_{i1}$ denote the elapsed time at the anchor visit. Since visit~1 is the anchor visit, $a_{i1}=0$ on the anchor-relative time scale. For a later visit~$t$, the age-adjusted anchor is:

\begin{equation}
\tilde{\mu}_{\mathrm{anchor},i}(a_{it}) = 
\mu_{\mathrm{post},i1} + 
\bigl[
\mu_{0,i}^{(\mathrm{adj})}(a_{it}) -
\mu_{0,i}^{(\mathrm{adj})}(a_{i1})
\bigr].
\label{eq:anchor_adj}
\end{equation}

\noindent This formulation preserves the individual's personalized anchor after the first observed score while allowing the reference level to follow expected age-related change over elapsed calendar time. Thus, the EBM-derived prior provides the initial cold-start reference, and the posterior after visit~1 provides the personal anchor for subsequent visits.
\subsection{Three-Tier Detection Rule}\label{note:three_tier_rule}

Cognitive decline was flagged using a three-tier rule that combines evidence of a negative trajectory with level-based checks against an age-adjusted anchor. Detection was active from Visit~1. At Visit~1, the EBM-derived baseline served as the cold-start anchor; from Visit~2 onward, the anchor was the posterior mean after incorporating the first observed score, providing a personalized comparison reference. Because slope estimation requires at least two observed scores, the slope signal contributes from Visit~2 onward, whereas level-based checks can be evaluated from Visit~1.

\paragraph{Below-anchor check.}
For binary flagging, the comparison reference was the age-adjusted anchor. At Visit~1, before a personal anchor was available, this reference was the age-adjusted EBM-derived baseline. From Visit~2 onward, the reference was the age-adjusted personal anchor, defined from the posterior mean after incorporating the first observed cognitive score. The age-adjusted anchor reference was therefore defined as:

\begin{equation}
\tilde{\mu}_{\mathrm{anchor},i}(a_{it}) =
\begin{cases}
\mu_{0,i}^{(\mathrm{adj})}(a_{it}), & t=1,\\[4pt]
\mu_{\mathrm{post},i1} +
\left[
\mu_{0,i}^{(\mathrm{adj})}(a_{it}) -
\mu_{0,i}^{(\mathrm{adj})}(a_{i1})
\right], & t \geq 2.
\end{cases}
\label{eq:anchor_piecewise}
\end{equation}
\noindent An individual was considered below anchor if either the posterior mean or the raw observed score fell below this reference by more than a margin $m$:

\begin{equation}
\mathrm{below\_anchor}_{it} =
\left(\mu_{\mathrm{post},it} < \tilde{\mu}_{\mathrm{anchor},i}(a_{it}) - m\right)
\;\lor\;
\left(y_{it} < \tilde{\mu}_{\mathrm{anchor},i}(a_{it}) - m\right).
\label{eq:below}
\end{equation}

\noindent The posterior mean provides a smoothed estimate of the individual's current cognitive level, whereas the raw observed score allows the rule to remain sensitive to abrupt drops that may not yet be fully reflected in the posterior mean.

\paragraph{Tier~0 --- Safety net (fixed, not tuned).}
Tier~0 served as a fixed safety-net rule for large level drops relative to the age-adjusted anchor. The standardized distance below the anchor was defined as

\begin{equation}
z_{\mathrm{below},it} =
\frac{\tilde{\mu}_{\mathrm{anchor},i}(a_{it}) - \mu_{\mathrm{post},it}}
{\sigma_{\mathrm{post},it}}.
\label{eq:z_below}
\end{equation}

\noindent An immediate flag was triggered if 
$z_{\mathrm{below},it} \geq z_{\mathrm{catch}}$ and 
$P(\mathrm{decline}_{it}) \geq 0.50$. This tier used a fixed threshold 
($z_{\mathrm{catch}}=2.0$) and was excluded from parameter optimization. When fewer than two observed scores were available, $P(\mathrm{decline}_{it})$ was set to 0.5, allowing Tier~0 to operate from Visit~1 onward based on the level-drop criterion alone.

\paragraph{Tier~1 --- Immediate rule (tuned).}
Tier~1 triggered an immediate flag when the posterior probability of decline exceeded the tuned immediate threshold and the individual was below the age-adjusted anchor:

\begin{equation}
\mathrm{Tier1}_{it} =
\left[P(\mathrm{decline}_{it}) \geq p_{\mathrm{imm}}\right]
\;\land\;
\mathrm{below\_anchor}_{it}.
\label{eq:tier1}
\end{equation}

\noindent This tier captures visits with strong evidence of decline together with a current level below the individualized reference.

\paragraph{Tier~2 --- Sustained rule (tuned).}
Tier~2 triggered a flag when a lower decline-probability threshold was met together with below-anchor status for $n_{\mathrm{confirm}}$ consecutive visits:

\begin{equation}
\mathrm{Tier2}_{it} =
\prod_{k=0}^{n_{\mathrm{confirm}}-1}
\mathds{1}\!\left[
P(\mathrm{decline}_{i,t-k}) \geq p_{\mathrm{sus}}
\;\land\;
\mathrm{below\_anchor}_{i,t-k}
\right] = 1.
\label{eq:tier2}
\end{equation}

\noindent This tier required both the sustained decline-probability threshold and the below-anchor condition to hold for $n_{\mathrm{confirm}}$ consecutive visits. The lower threshold $p_{\mathrm{sus}} < p_{\mathrm{imm}}$ allowed gradual decline to be detected, while the consecutive-visit requirement reduced sensitivity to transient fluctuations.

\paragraph{Final flag decision.}
At each visit, the final binary flag was defined as

\begin{equation}
\mathrm{flag}_{it} =
\mathrm{Tier0}_{it}
\;\lor\;
\mathrm{Tier1}_{it}
\;\lor\;
\mathrm{Tier2}_{it}.
\label{eq:flag}
\end{equation}

\paragraph{Confirmation and flag stability.}
For individual-level summaries, flag status was evaluated over the longitudinal sequence rather than from a single isolated observation. Once an alert episode was triggered, subsequent visits were evaluated using the recovery rules described below.

\paragraph{Recovery (Unflagging) Mechanism}

After a decline flag was activated, subsequent visits were evaluated to determine whether the flag should remain active or be cleared. Recovery rules were applied differently in HRS and ADNI because the two cohorts used different outcome definitions and showed different rates of apparent reversion.

\paragraph{HRS.}
In HRS, unflagging was permitted when subsequent visits no longer satisfied any of the three detection tiers. The number of required clear visits depended on the tier that initiated the alert episode:

\begin{itemize}
    \item Tier~0/1 alerts, which reflect single-visit signals, required 1 clear visit to unflag.
    \item Tier~2 alerts, which reflect sustained decline, required $n_{\mathrm{recover}}$ consecutive clear visits to unflag.
\end{itemize}

\noindent The recovery parameter was fixed at $n_{\mathrm{recover}}=2$ and was not optimized. If an individual was unflagged and later re-flagged, the second alert episode was treated as permanent.

\paragraph{ADNI.}
In ADNI, unflagging was disabled. Because ADNI outcomes are based on clinician-assigned diagnostic status, apparent reversions were rare and unflagging was empirically less beneficial.

\subsection{Ground Truth Definition}\label{note:ground_truth}

For post-hoc evaluation, the primary endpoint was sustained cognitive decline, defined from the observed diagnostic or classification trajectory. For individual~$i$, sustained decline was defined as:

\begin{equation}
\mathrm{sustained\_decline}_i =
\bigl(n_{\mathrm{worse},i} \geq 2\bigr)
\;\lor\;
\bigl(C_{i,\mathrm{last}} \succ C_{i,\mathrm{first}}\bigr),
\label{eq:gt}
\end{equation}

\noindent where $C_{i,\mathrm{first}}$ and $C_{i,\mathrm{last}}$ denote the individual's first and last observed diagnostic/classification states, $n_{\mathrm{worse},i}$ is the number of visits at which the classification was worse than $C_{i,\mathrm{first}}$, and $\succ$ denotes a higher ordinal rank on the classification scale. This definition captures individuals with either at least two visits in a worse state than baseline, indicating repeated or sustained worsening, or a final classification worse than baseline, indicating terminal deterioration even when worsening is observed only at the last visit. Single transient worsening episodes were not counted as sustained decline unless the individual ended in a worse state.

\paragraph{Classification scales}
\begin{itemize}
    \item \textbf{HRS:} Langa--Weir classification derived from the HRS 27-point cognitive summary score: Normal ($\geq 12$), CIND ($7$--$11$), and Dementia ($0$--$6$). Ordinal mapping: Normal\,=\,0, CIND\,=\,1, Dementia\,=\,2.
    \item \textbf{ADNI:} Clinician-assigned diagnosis: CN (Cognitively Normal)\,=\,0, MCI (Mild Cognitive Impairment)\,=\,1, Dementia\,=\,2.
\end{itemize}

\noindent These ground-truth labels were not accessed by the detection framework during prediction. They were used only for post-hoc performance evaluation.

\subsection{Forecasting Calibration and Detection-Rule Optimization}\label{note:forecasting_detection_optimization}

This step involved two distinct validation-set procedures. First, the temporal decay rate $\lambda$ was calibrated to improve one-step-ahead forecasting of cognitive scores. Second, the detection-rule parameters were optimized to identify sustained cognitive decline using the ground-truth endpoint defined above. Thus, $\lambda$ was selected using a forecasting objective, whereas the detection thresholds were selected using a hierarchical classification objective. Test data were held out until final evaluation.

\subsubsection{Temporal Decay Calibration for Forecasting}\label{note:lambda}

The decay rate $\lambda$ controls how rapidly older observations are down-weighted in the Bayesian level update (Eq.~\ref{eq:decay}). Because temporal distances were measured in years, $\lambda$ is interpreted on a per-year scale. The decay rate was selected independently of the binary detection thresholds by minimizing one-step-ahead mean absolute error (MAE) on the validation set:

\begin{equation}
\lambda^* =
\mathop{\mathrm{arg\,min}}_{\lambda \in \Lambda}
\frac{1}{N_{\mathrm{pred}}}
\sum_{(i,t)\in\mathcal{V}}
\left|y_{it} - \mu_{\mathrm{post},i,t-1}^{(\lambda)}\right|,
\label{eq:lambda_sel}
\end{equation}

\noindent where $\mathcal{V}$ is the set of validation-set one-step-ahead prediction pairs with at least one prior observed visit, $N_{\mathrm{pred}} = |\mathcal{V}|$, $\mu_{\mathrm{post},i,t-1}^{(\lambda)}$ is the posterior mean for individual~$i$ after visit~$t-1$ computed using decay rate $\lambda$, and $y_{it}$ is the observed cognitive score at visit~$t$. The candidate grid used in the present experiments was  $\Lambda = \{0.05, 0.10, 0.15, 0.20, 0.30, 0.50, 0.70, 0.80, 1.00\}$.

\subsubsection{Detection-Rule Optimization}\label{note:det_rule_optimization}

After selecting the temporal decay rate $\lambda$, the binary detection-rule parameters were optimized on the validation data to identify individuals who meet the sustained cognitive decline endpoint (Eq.~\ref{eq:gt}). This optimization was separate from the one-step-ahead forecasting calibration described above. It tuned the Tier~1 immediate-decline threshold $p_{\mathrm{imm}}$, the Tier~2 sustained-decline threshold $p_{\mathrm{sus}}$, and the number of consecutive visits required for Tier~2 confirmation, $n_{\mathrm{confirm}}$.

\noindent The following grids define the candidate ranges used for validation-based calibration in the present experiments; they are not intended as fixed thresholds for other datasets or deployment settings.

\begin{table}[!ht]
\centering
\caption{Grid search ranges for detection-rule optimization.}
\label{tab:grids}
\small
\begin{tabular}{p{0.10\textwidth} p{0.47\textwidth} p{0.34\textwidth}}
\hline
\textbf{Parameter} & \textbf{Grid values} & \textbf{Description} \\
\hline
$p_{\mathrm{imm}}$ 
& $\{0.55, 0.60, 0.65, 0.70, 0.75, 0.80, 0.85, 0.90, 0.95\}$ 
& Tier~1 immediate-decline threshold \\

$p_{\mathrm{sus}}$ 
& $\{0.40, 0.45, 0.50, 0.55, 0.60, 0.65, 0.70\}$ 
& Tier~2 sustained-decline threshold \\

$n_{\mathrm{confirm}}$ 
& $\{2, 3, 4\}$ 
& Consecutive visits required for Tier~2 confirmation \\
\hline
\end{tabular}
\end{table}

\paragraph{Hierarchy constraint.}
To preserve the intended hierarchy between the immediate and sustained rules, valid configurations were required to satisfy:

\begin{equation}
p_{\mathrm{imm}} \;\geq\; p_{\mathrm{sus}} + \Delta_{\min},
\label{eq:gap_constraint}
\end{equation}

\noindent where $\Delta_{\min}$ is the minimum threshold gap, specified by age band (Table~\ref{tab:age_params}). This constraint ensures that Tier~1 requires stronger instantaneous evidence than Tier~2, whereas Tier~2 can use a lower probability threshold because it also requires persistence across visits.

\paragraph{Fixed parameters.}
The following parameters were fixed a priori and were not tuned during detection-rule optimization:

\begin{itemize}
    \item $z_{\mathrm{catch}} = 2.0$ --- Tier~0 standardized below-anchor threshold.
    \item $n_{\mathrm{recover}} = 2$ --- number of consecutive clear visits required to unflag a Tier~2 alert in HRS.
    \item $r = 0.70$ --- test--retest reliability coefficient used to compute $\mathrm{SEM}$.
    \item Anchor index $=1$ --- the first observed cognitive score establishes the personal anchor after EBM prior initialization.
\end{itemize}

\paragraph{Hierarchical detection objective.}
Among all valid detection-rule configurations satisfying Eq.~\ref{eq:gap_constraint}, the selected configuration was determined using a four-step hierarchical objective. This objective was designed to prioritize early identification of sustained cognitive decline while maintaining acceptable specificity among individuals with no evidence of diagnostic worsening.

\begin{enumerate}
\item \textbf{Hard constraint: specificity floor.} 
Configurations were first filtered according to specificity among hard negatives. Hard negatives were defined as individuals who never had a visit with a worse diagnostic/classification state than their baseline state. Only configurations with hard-negative specificity $\geq 0.70$ were retained. If no configuration satisfied this floor, the constraint was relaxed to within 0.05 of the best achievable hard-negative specificity.

\item \textbf{Primary objective: early detection reward.} 
Among configurations satisfying the specificity constraint, we maximized a saturating early-detection reward:

\begin{equation}
R = \frac{1}{n_{\mathrm{pos}}}
\sum_{i \in \mathcal{P}}
\mathds{1}\!\left(\ell_i > 0\right)
\left[1 - \exp\!\left(-k \ell_i\right)\right],
\label{eq:early_reward}
\end{equation}

\noindent where $\mathcal{P}$ is the set of validation-set positives, $n_{\mathrm{pos}} = |\mathcal{P}|$, $\ell_i$ is the lead time in visits between the first flag and the first observed transition to a worse diagnostic/classification state, and $k$ is a saturation rate. Positive cases not flagged before transition contributed zero reward. Configurations within $\epsilon_R$ of the maximum reward were retained for the next step.

\item \textbf{Secondary objective: $F_2$ score.} 
Among configurations tied on the early-detection reward, we maximized the $F_2$ score, a recall-weighted harmonic mean of precision and recall:

\begin{equation}
F_2 = \frac{5 \cdot \mathrm{precision} \cdot \mathrm{recall}}
{4 \cdot \mathrm{precision} + \mathrm{recall}}.
\label{eq:f2}
\end{equation}

\noindent Configurations within $\epsilon_{F_2}$ of the maximum $F_2$ score were retained.

\item \textbf{Tie-break: visit-level false-positive burden.} 
Among the remaining configurations, the selected configuration was the one with the lowest visit-level false-positive burden among hard negatives, defined as the fraction of hard-negative visits that were flagged.
\end{enumerate}

\subsubsection{Age-Stratified Detection-Rule Optimization}\label{note:age_stratified}

After global detection-rule optimization, the detection thresholds were refined independently within four age bands to account for age-related differences in cognitive variability, decline prevalence, and acceptable sensitivity--specificity trade-offs. The same hierarchical detection objective described in Supplementary Note~\ref{note:det_rule_optimization} was applied separately to each age band's validation subset. The temporal decay rate $\lambda$ was not reselected within age bands; age-stratified optimization applied only to the binary detection-rule parameters $p_{\mathrm{imm}}$, $p_{\mathrm{sus}}$, and $n_{\mathrm{confirm}}$.

\noindent The margin $m$ and the minimum threshold gap $\Delta_{\min}$ were specified a priori by age band and were not optimized:

\begin{table}[h]
\centering
\caption{Pre-specified age-band-specific parameters used during detection-rule optimization.}
\label{tab:age_params}
\small
\begin{tabularx}{\textwidth}{l c c X}
\hline
\textbf{Age band} & \textbf{Margin $m$} & \textbf{Min gap $\Delta_{\min}$} & \textbf{Rationale} \\
\hline
50--64 & 0.5 & 0.20 & Buffer against test--retest noise; gradual decline expected to require stronger persistence evidence. \\
65--74 & 0.5 & 0.20 & Buffer against test--retest noise; gradual decline expected to require stronger persistence evidence. \\
75--84 & 0.0 & 0.10 & Small deviations from the anchor may be clinically meaningful; abrupt decline is more common. \\
85+    & 0.0 & 0.10 & Small deviations from the anchor may be clinically meaningful; abrupt decline is more common. \\
\hline
\end{tabularx}
\end{table}

\noindent The margin $m$ is measured in cognitive-score points and is subtracted from the age-adjusted anchor in the below-anchor check (Eq.~\ref{eq:below}). For participants aged 50--74, a 0.5-point margin provides a buffer against normal test--retest variability. For participants aged 75 and older, the margin is set to zero so that smaller deviations from the age-adjusted anchor can contribute to flagging. The minimum gap $\Delta_{\min}$ enforces separation between the Tier~1 immediate threshold $p_{\mathrm{imm}}$ and the Tier~2 sustained threshold $p_{\mathrm{sus}}$ (Eq.~\ref{eq:gap_constraint}). The wider gap in younger age bands gives Tier~2 a broader probability range for detecting gradual decline, whereas the narrower gap in older age bands allows more similar immediate and sustained thresholds when abrupt decline is more common.
\subsection{Cohort-Specific Implementation Details}

\subsubsection{HRS (Health and Retirement Study)}\label{note:HRS}

\paragraph{Data source and outcome.}
HRS analyses used the RAND HRS Longitudinal File 2022 v1, Waves~3--15 (1996--2020). The cognitive outcome was Total Word Recall, using RwTR20 for Waves~3--13 and RwTR20P for Waves~14--15, with scores ranging from 0 to 20.

\paragraph{Baseline features.}
The EBM baseline model used 24 Visit~1 features spanning demographics, health status, medical conditions, mental health and functional status, lifestyle, socioeconomic status, and healthcare utilization. Household income was transformed as $\mathrm{sign}(x)\log(1+|x|)$.

\paragraph{Inclusion/exclusion criteria.}
Participants were included if they were in-person respondents, had a valid word recall score, had at least two cognition waves, were aged $\geq 50$ at the first visit, were not classified as Dementia at baseline according to the Langa--Weir classification, and had at least 60\% feature completeness at Visit~1. Features with less than 50\% availability at Visit~1 were excluded.

\paragraph{Train/validation/test split.}
Data were split at the participant level using \texttt{GroupShuffleSplit} with a 70\%/15\%/15\% train/validation/test split and \texttt{random\_state\,=\,42}. All waves from the same participant were kept in the same split. The EBM was trained on Visit~1 data from the training set only; forecasting calibration and detection-rule optimization were performed on the validation set; final performance was evaluated on the held-out test set.

\paragraph{Key estimated quantities.}
For HRS, the EBM residual standard deviation was $\sigma_0=2.860$, the standard error of measurement was $\mathrm{SEM}=1.784$ using $\mathrm{SD}_y=3.256$ and reliability $r=0.70$, the empirical slope prior standard deviation was $\sigma_\beta=0.509$ points/year, and the selected temporal decay rate was $\lambda^*=0.20$.

\subsubsection*{ADNI (Alzheimer's Disease Neuroimaging Initiative)}\label{note:ADNI}

\paragraph{Data source and outcome.}
ADNI analyses used longitudinal data from ADNI-1 through ADNI-3. The cognitive outcome was AVLT Delayed Recall, with scores ranging from 0 to 15.

\paragraph{Baseline features.}
The EBM baseline model used eight baseline features: age, sex, education years, race, partnered status, Functional Assessment Questionnaire (FAQ) total, Geriatric Depression Scale (GDS) total, and BMI.

\paragraph{Inclusion/exclusion criteria.}
Participants were included if they had a valid AVLT delayed recall score, a valid diagnosis, at least two visits, and were not diagnosed with Dementia at baseline. Features with less than 50\% availability were excluded.

\paragraph{Validation strategy.}
ADNI was evaluated using 5-fold stratified cross-validation at the individual level, stratified by first diagnosis (CN vs.\ MCI), with \texttt{random\_state\,=\,42}. For each fold $k$, fold $k$ was used as the test fold, fold $(k+1)\bmod 5$ as the validation fold, and the remaining three folds as training data. Lambda selection and detection-rule optimization were performed within the validation fold, and held-out test-fold predictions were pooled across the five folds for final evaluation.

\paragraph{Key estimated quantities.}
Across the five ADNI folds, the EBM residual standard deviation was $\sigma_0=2.974\pm0.018$, the standard error of measurement was $\mathrm{SEM}=1.815\pm0.009$, and the selected temporal decay rate was $\lambda^*=0.96\pm0.08$. Unflagging was disabled in ADNI.

\subsection{Validation-Calibrated Detection Parameters}

The Tier~1 threshold $p_{\mathrm{imm}}$, Tier~2 threshold $p_{\mathrm{sus}}$, and Tier~2 confirmation length $n_{\mathrm{confirm}}$ were treated as validation-calibrated implementation parameters rather than fixed components of the framework. In other applications, these thresholds can be recalibrated according to the target population, cognitive instrument, follow-up frequency, and desired sensitivity--specificity trade-off. In the present experiments, age-band-specific values were selected on validation data using the hierarchical detection objective described above and then applied unchanged to held-out test data.

\subsection{Evaluation Metrics}\label{note:evaluation_metrics}

\paragraph{Composite decline score for discrimination analyses.}
For ROC/AUC-based discrimination analyses, we computed a continuous composite decline score. This score was used only for post-hoc evaluation and did not enter the binary three-tier flagging rules. It combines the posterior probability of a negative slope with the posterior probability that the current cognitive level is below the age-adjusted EBM-derived baseline:

\begin{align}
P(\mathrm{below\ EBM}_{it}) 
&= \Phi\!\left(
\frac{\mu_{0,i}^{(\mathrm{adj})}(a_{it}) - \mu_{\mathrm{post},it}}
{\sigma_{\mathrm{post},it}}
\right),
\label{eq:p_below}\\[4pt]
\mathrm{decline\_score}_{it} 
&= P(\mathrm{decline}_{it}) \times P(\mathrm{below\ EBM}_{it}).
\label{eq:composite}
\end{align}

\noindent Here, $P(\mathrm{decline}_{it})$ is the posterior probability that the slope is below zero (Eq.~\ref{eq:p_decline}), and $P(\mathrm{below\ EBM}_{it})$ is the posterior probability that the current latent cognitive level is below the age-adjusted EBM-derived baseline.

\paragraph{Individual-level detection.}
An individual was classified as declining if their visit-level flag remained active at their last observed visit. Individuals flagged only once who subsequently met the recovery criteria were not identified as declining. However, individuals who were flagged, unflagged, and then re-flagged (a second flagging episode) retained a permanent flag regardless of subsequent recovery, reflecting persistent evidence of decline. Precision, recall, $F_1$, $F_2$, and specificity were calculated against the sustained decline endpoint (Eq.~\ref{eq:gt}).

\paragraph{Early detection.}
Among true positives, early detection was summarized as the proportion flagged before the first diagnostic transition visit and the corresponding lead time, measured in visits and years.

\paragraph{One-step-ahead forecast accuracy.}
For visits $t \geq 2$, PRISM's posterior mean after visit $t-1$ was compared with the observed cognitive score at visit~$t$. Forecasting performance was summarized using mean absolute error (MAE). PRISM was compared with two reference forecasting baselines. First, the age--education--sex adjusted population mean assigned each participant the training-set mean score for their demographic stratum:

\begin{equation}
\hat{y}_{it}^{(\mathrm{norm})}
=
\bar{y}_{\mathrm{train}}
\left(
\mathrm{age\ band}_{it},
\mathrm{education\ band}_{i},
\mathrm{sex}_{i}
\right).
\label{eq:population_norm}
\end{equation}

\noindent Age bands were 50--64, 65--74, 75--84, and 85+, and education bands were $<12$, 12--15, and 16+ years. Second, the cumulative individual average used the mean of all prior observed scores for that participant:

\begin{equation}
\hat{y}_{it}^{(\mathrm{cum})} = \frac{1}{t-1}\sum_{j=1}^{t-1} y_{ij}.
\label{eq:cumulative_average}
\end{equation}

\noindent The cumulative average was available from visit~2 onward and provided a purely individual-history baseline without population-level initialization.

\paragraph{AUC analyses.}
Individual-level AUC was computed using the maximum posterior decline probability across visits and the maximum composite decline score across visits as continuous predictors of the sustained decline endpoint. Visit-level AUC was computed using the composite decline score as a continuous predictor of whether the individual's diagnostic/classification state at a given visit was worse than their baseline state. For comparison with a standard repeated-measures model, we also fit a linear mixed-effects model with fixed effects for visit, age, education, and sex, and individual-specific random intercepts and slopes for visit:

\begin{equation}
y_{ij} =
\beta_0 + \beta_1 \mathrm{visit}_{ij}
+ \beta_2 \mathrm{age}_{ij}
+ \beta_3 \mathrm{education}_{i}
+ \beta_4 \mathrm{sex}_{i}
+ u_{0i} + u_{1i}\mathrm{visit}_{ij}
+ \varepsilon_{ij}.
\label{eq:lme}
\end{equation}

\noindent The LME comparison used the individual-specific slope estimate,
$\hat{\beta}_1 + \hat{u}_{1i}$, as the continuous decline signal, with more negative slopes indicating a greater decline risk.
\subsection{Software and Reproducibility}

All analyses were implemented in Python~3 using NumPy, pandas, SciPy, scikit-learn, and InterpretML. Random seeds were fixed where applicable (\texttt{random\_state\,=\,42}, \texttt{np.random.seed(42)}). EBM models were trained using \texttt{ExplainableBoostingRegressor} from InterpretML. 

\noindent Bayesian updating, slope estimation, and the three-tier detection logic were implemented directly in Python without external Bayesian inference libraries.
